\documentclass[a4paper,11pt]{article}
\usepackage[utf8]{inputenc}
\usepackage{amsmath,amssymb,amsthm,mathrsfs,amsfonts,dsfont,bm} 
\usepackage[dvipsnames,table,xcdraw]{xcolor}
\usepackage{array,graphicx}
\usepackage{hyperref}
\usepackage{float}
\usepackage[margin=2.5cm]{geometry}
\usepackage[toc]{appendix}
\usepackage{enumerate}
\usepackage{multirow}
\usepackage{tikz}
\usepackage[round, sort]{natbib}
\usepackage{setspace}
\usepackage{lineno}

\newtheorem{theorem}{Theorem}
\newtheorem{lemma}{Lemma}

\newcommand{\modif}[1]{\textcolor{black}{#1}}

\DeclareMathOperator*{\Esp}{\mathbb{E}}
\DeclareMathOperator*{\Var}{\mathbb{V}}
\DeclareMathOperator*{\Cov}{\mathbb{C}\text{ov}}
\DeclareMathOperator*{\prob}{\mathds{P}}

\newcommand\jk{{jk}}

\newcommand\Ncal{\mathcal{N}}
\newcommand\Pcal{\mathcal{P}}
\newcommand\Tcal{\mathcal{T}}
\newcommand\mt{\widetilde{m}}
\newcommand\St{\widetilde{S}}
\newcommand\mbt{\widetilde{\bf m}}
\newcommand\Sbt{\widetilde{\bf S}}

\newcommand\gammab{{\boldsymbol{\gamma}}}
\newcommand\betab{{\boldsymbol{\beta}}}
\newcommand\thetab{{\boldsymbol{\theta}}}
\newcommand\Sigmab{{\boldsymbol{\Sigma}}}
\newcommand\cst{\text{cst}}

\newcommand\Hb{{\bf H}}
\newcommand\Mb{{\bf M}}
\newcommand\Qb{{\bf Q}}
\newcommand\Wb{{\bf W}}
\newcommand\Xb{{\bf X}}
\newcommand\xb{{\bf x}}
\newcommand\Yb{{\bf Y}}
\newcommand\Zb{{\bf Z}}

\newcommand\pt{\widetilde{p}}

\newcommand{\nodesize}{1em}
\newcommand{\edgeunit}{4*\nodesize}
\tikzstyle{covariate}=[draw, rectangle, minimum width=\nodesize, minimum height=\nodesize, inner sep=0, color=black]
\tikzstyle{covmiss}=[draw, minimum width=\nodesize, minimum height=\nodesize, inner sep=0, color=gray, text=gray]
\tikzstyle{observed}=[draw, circle, minimum width=\nodesize, inner sep=0, color=black]
\tikzstyle{edge}=[-, line width=1pt, color=black]
\tikzstyle{edgemiss}=[-, line width=1pt, dashed, color=gray]

\usepackage[affil-it]{authblk}

\title{\textbf{
\modif{Tree-based Inference of Species Interaction Network from Abundance Data}
}}

\author{Raphaëlle Momal$^1$%
  \thanks{Electronic address: \texttt{raphaelle.momal@agroparistech.fr}; Corresponding author}, \hspace{0.3cm} Stéphane Robin$^1$, \hspace{0.3cm} Christophe Ambroise$^2$}
\affil{1: UMR MIA-Paris, AgroParisTech, INRA, Université Paris-Saclay, 75005 Paris, France\\
2: Laboratoire de Mathématiques et Modélisation d'Évry, 23 bvd de France, Évry, France}

\date{Dated: \today}

\date{\today}

\begin{document}
\setlength{\parindent}{0ex}
\maketitle

% \tableofcontents \bigskip
\newpage
%%%%%%%%%%%%%%%%%%%%%%%%%%%%%%%%%%%%%%%%%%%%%
\begin{abstract}
%Point 1: set the context for and purpose of the work;
%Point 2: indicate the approach and methods;
%Point 3: outline the main results;
%Point 4: identify the conclusions and the wider implications.
%\modif{A REPRENDRE}
\noindent 1. The behavior of ecological systems mainly relies on the interactions between the species it involves. 
%In many situations, these interactions are not observed and have to be inferred from species abundance data. 
\modif{We consider the problem of inferring the species interaction network from abundance data.}
To be relevant, any  network \modif{inference} methodology needs to handle count data and to account for possible environmental effects. It also needs to distinguish between direct interactions and indirect associations and graphical models provide a convenient framework for this purpose. \\
2. We introduce a generic statistical model for network \modif{inference} based on abundance data. The model includes fixed effects to account for environmental covariates and sampling efforts, and correlated random effects to encode species interactions. The inferred network is obtained by averaging over all possible tree-shaped (and therefore sparse) networks, in a computationally efficient manner. An output of the procedure is the probability for each edge to be part of the underlying network.  \\
3. A simulation study shows that the proposed methodology compares well with state-of-the-art approaches, even when the underlying \modif{graph} strongly differs from a tree. The analysis of two datasets highlights the influence of covariates on the inferred network. \\
4. Accounting for covariates is critical to avoid spurious edges. 
The proposed approach could be extended to perform network comparison or to look for missing species.

\modif{species interaction network}
\end{abstract}

% \section*{Notations} \input{notations}
\newpage
\section{Introduction}

\modif{There is a growing awareness of biotic interactions being crucial components of biodiversity and relevant descriptors of ecosystems \citep{valiente,jordanoS}. 
Such interactions can be conveniently represented by networks, which have been increasingly studied and used in recent years for describing and understanding living systems in ecology \citep{poisot}, microbiology \citep{faust} or genomics \citep{evans}. 
Observing species interactions is a laborious task which restricts them to certain categories (e.g. trophic, pollination), while many other mutualistic and/or antagonistic interactions may be hard to observe and key in the system organization (e.g. communication, shelter sharing). Many efforts have been devoted in the last decade to get a more complete picture of the biotic interactions existing between  species living in the same niche.}

\modif{\paragraph{Network reconstruction.} 
A first attempt consists in using observed interactions  to predict other possible links based on species traits matching \citep[see e.g.][]{OlF15,BGT16,WeG17,GrW18}. The  interaction strength can also be predicted \citep{WeO13}. This  can be viewed as a prediction task, and modern approaches arising from signal processing and machine learning have been also proposed \citep{DPL17,SPW17,DPD17}. We name these approaches {\sl network reconstruction} to distinguish them from {\sl network inference}, which is the problem we consider in this article.}

\begin{figure}[H]
    \centering
    \begin{tabular}{ccccc}
%        $\Xb$ & & $\Yb$ & & $\widehat{G}$ \\
        {\tt \begin{tabular}{rr}
date & site \\
%\hline
apr93 & km03 \\
apr93 & km03 \\
apr93 & km03 \\
apr93 & km03 \\
apr93 & km17 \\
apr93 & km17 \\
\vdots & \vdots
\end{tabular} } & \qquad &
        {\tt \begin{tabular}{rrrrr}
EFI & ELA & GDE & GME & HFA \\
%\hline
 71 &   1 &   5 &   6 &   0 \\
118 &   2 &   3 &   0 &   0 \\
 69 &   0 &   6 &   2 &   0 \\
 56 &   0 &   0 &   0 &   0 \\
  0 &   1 &   1 &   0 &   0 \\
  0 &   0 &   2 &   0 &   0 \\
\vdots & \vdots & \vdots & \vdots & \vdots
\end{tabular}
%
%\begin{tabular}{rrrrrrr}
%\dots & EFI & ELA & GDE & GME & HFA & \dots \\
%\hline
%&  71 &   1 &   5 &   6 &   0 & \\
%& 118 &   2 &   3 &   0 &   0 & \\
%&  69 &   0 &   6 &   2 &   0 & \\
%&  56 &   0 &   0 &   0 &   0 & \\
%&   0 &   1 &   1 &   0 &   0 & \\
%&   0 &   0 &   2 &   0 &   0 & \\
%& \vdots & \vdots & \vdots & \vdots & \vdots
%\end{tabular}
 } & \qquad &
        \begin{tabular}{c}
        \includegraphics[width=.3\linewidth]{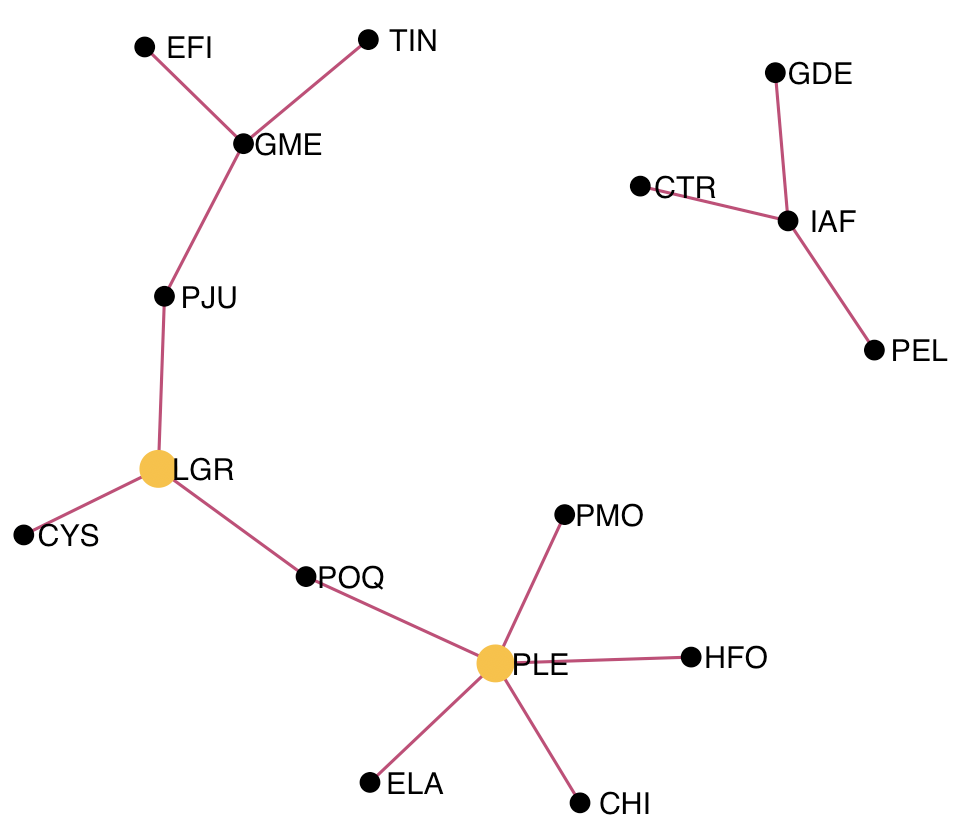}
        \end{tabular} \\
        ($a$) covariates & & ($b$) species abundances & & ($c$) inferred network \\
    \end{tabular}
    \caption{Aim of species interaction network inference from abundance data. Data sample from the Fatala river dataset (see Section \ref{sec:datasets}). }
    \label{fig:networkinference}
\end{figure}

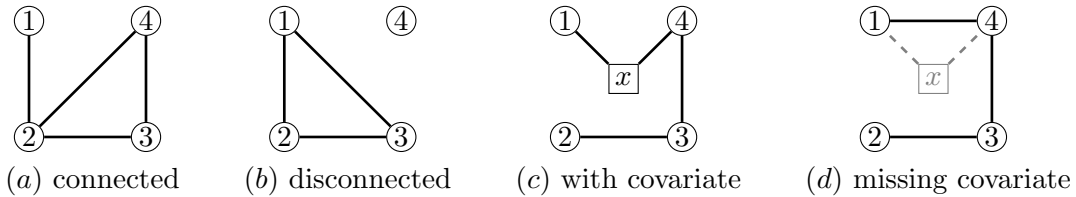
\begin{figure}[H]
    \centering
    \begin{tabular}{ccccccc}
          \begin{tikzpicture}
  \node[observed] (1) at (-0.5*\edgeunit,  .5*\edgeunit) {$1$};
  \node[observed] (2) at (-0.5*\edgeunit, -.5*\edgeunit) {$2$};
  \node[observed] (3) at ( 0.5*\edgeunit, -.5*\edgeunit) {$3$};
  \node[observed] (4) at ( 0.5*\edgeunit,  .5*\edgeunit) {$4$};
  \draw[edge] (1) to (2); \draw[edge] (2) to (3); \draw[edge] (2) to (4); \draw[edge] (3) to (4); 
  \end{tikzpicture} & \qquad &
          \begin{tikzpicture}
  \node[observed] (1) at (-0.5*\edgeunit,  .5*\edgeunit) {$1$};
  \node[observed] (2) at (-0.5*\edgeunit, -.5*\edgeunit) {$2$};
  \node[observed] (3) at ( 0.5*\edgeunit, -.5*\edgeunit) {$3$};
  \node[observed] (4) at ( 0.5*\edgeunit,  .5*\edgeunit) {$4$};
  \draw[edge] (1) to (2); \draw[edge] (2) to (3);  \draw[edge] (1) to (3); 
  \end{tikzpicture} & \qquad &
          \begin{tikzpicture}
  \node[observed] (1) at (-0.5*\edgeunit,  .5*\edgeunit) {$1$};
  \node[observed] (2) at (-0.5*\edgeunit, -.5*\edgeunit) {$2$};
  \node[observed] (3) at ( 0.5*\edgeunit, -.5*\edgeunit) {$3$};
  \node[observed] (4) at ( 0.5*\edgeunit,  .5*\edgeunit) {$4$};
  \node[covariate] (x) at (0.0*\edgeunit,  .0*\edgeunit) {$x$};
  \draw[edge] (2) to (3); \draw[edge] (3) to (4); 
  \draw[edge] (x) to (1); \draw[edge] (x) to (4);
  \end{tikzpicture} & \qquad &
          \begin{tikzpicture}
  \node[observed] (1) at (-0.5*\edgeunit,  .5*\edgeunit) {$1$};
  \node[observed] (2) at (-0.5*\edgeunit, -.5*\edgeunit) {$2$};
  \node[observed] (3) at ( 0.5*\edgeunit, -.5*\edgeunit) {$3$};
  \node[observed] (4) at ( 0.5*\edgeunit,  .5*\edgeunit) {$4$};
  \node[covmiss] (x) at (0.0*\edgeunit,  .0*\edgeunit) {$x$};
  \draw[edge] (2) to (3); \draw[edge] (3) to (4); 
  \draw[edgemiss] (x) to (1); \draw[edgemiss] (x) to (4);
  \draw[edge] (1) to (4); 
  \end{tikzpicture} \\
        ($a$) connected & & ($b$) disconnected & & ($c$) with covariate & & ($d$) missing covariate 
    \end{tabular}
    \caption{Examples of graphical models.}
    \label{fig:graphmodel}
\end{figure}

\modif{\paragraph{Network inference.} 
Network inference approaches also aim at retrieving the interactions among species, but do not rely on observed interactions and therefore, remain agnostic as for their type. Such approaches have been developed in many domains ranging from cell biology \citep[][to infer gene regulatory networks]{Fri04} to neurosciences \citep[][to deciphere brain connectivity structures]{ZhC18}. In ecology, it will typically aim at inferring the set of biotic interactions linking species from the same guild.  As  summarized in Fig.~\ref{fig:networkinference}, network inference takes as input measures on species at similar sites, and returns a network of species direct interactions. The importance of distinguishing between direct interaction and indirect association between species is explained in \citet{PWT19}. To be accurate, network inference must account for environmental covariates to prevent the inference of spurious interactions resulting from abiotic effects. Fig.~\ref{fig:graphmodel} illustrates this phenomenon: ($c$) corresponds to the case where two species (1 and 4) are not in direct interaction, but are affected by the variations of the same environmental covariate $x$. ($d$) displays the network when $x$ is not accounted for: a spurious edge appears between these two species.}

\modif{\paragraph{Joint species distribution models.} The rational between network inference is that interactions between species must affect their joint distribution in a series of similar sites. 
Such approaches necessarily rely on a {\sl joint} species distribution model (JSDM), as opposed to species distribution models \citep{SDM} where species are traditionally considered as disconnected entities. 
A JSDM is a probabilistic model describing the species simultaneous presence/absence \citep{Har15,OTN17} or joint abundances \citep{PHW18,PWT19}. An important feature of JSDMs is to include environmental covariates to account for abiotic interactions. \\
Recently, latent variable models have received attention in community ecology as they provide a convenient way to model the dependence structure between species  \citep{WBO15}. The JSDM proposed by \cite{PHW18,PWT19} involves a latent layer. So does the Poisson log-Normal model \citep[PLN,][]{AiH89}, which combines generalized linear models to account for covariates and offsets, and a Gaussian latent structure to describe the species interactions. It can be seen as a multivariate mixed model, in which correlated random effects encode the dependency between the species abundances. }

\modif{\paragraph{Graphical models: a generic framework for network inference.}
Although they describe the dependence structure between the distributions of all the species from a same niche, JSDM are not sufficient to perform network inference as they do not distinguish  indirect associations from direct interactions \citep{DBD18}. Graphical models \citep{Lau96} provide a probabilistic framework to do so and, in the same time, a formal definition of the network to be inferred. This formalism is therefore especially appealing for the inference of species interaction networks \citep{PWT19}.
In an undirected graphical model \citep[which is the same as a Markov random field:][]{CWL18}, two species are connected if they are {\sl dependent} conditional on all other species, that is if the variations of their respective abundances would still be  correlated if ever both the environmental conditions and the abundances of all other species were kept fixed. Two species are unconnected if they are {\sl independent} conditional on all other species: the observed correlation between them only results from a series of links with other species \citep{morueta2016network}. 
Fig.~\ref{fig:graphmodel} illustrates the concept of conditional dependence/independence with toy graphical models. In ($a$), the network is connected so all species are interdependent: an association exists between any two of them. However, 1 is only directly interacting with 2 which mediates its association with 3 and 4: 1 is independent from them conditional on 2.
%does not have direct interactions with  3 and 4: its association with them is mediated by  2, and it is independent from them conditionally on 2. 
In ($b$), the network is disconnected: species 4 is independent from all others. This illustrates that graphical models enjoy all the desirable properties to represent interactions between species in an interpretable manner, so that they can be used as the mathematical counterpart of species interaction networks.}

\modif{\paragraph{Network inference: the general problem.}
Network inference methods attempts to retrieve the graphical model underlying the distribution of observed data. In every domains, network inference is impeded by the huge number of possible graphs for a given set of nodes, which increases super-exponentially with the latter (more than $10^{13}$ undirected graphs can be drawn between 10 nodes, and more than $10^{57}$ between 20). The exploration of the graph space is therefore intractable from a combinatorial point of view. To reduce the search space, a common and reasonable assumption is that a relatively small fraction of species pairs are in direct interaction: the network is sparse. In the case of continuous observations, one of the most popular approach is the graphical lasso \citep[glasso:][]{GLasso} which takes advantage of the properties of Gaussian graphical models (GGM) to efficiently infer a sparse network. %In the Gaussian setting, correlations are related to direct or indirect associations, whereas partial correlations are specific to direct interactions. \\
Alternatively, tree-based approaches have been proposed: \cite{ChowLiu} first made the too stringent assumption that the network is made of a single spanning tree (that is connecting all nodes without any loop, as in Fig.~\ref{fig:treeaveraging}).
%, which also rely on efficient algebraic tools. Also early approaches \citep{ChowLiu} made the much too stringent assumption that the network is made of a single spanning tree (i.e. has no loop), 
More recent approaches introduced by \cite{,MeilaJaak} and \cite{kirshner} rely on efficient algebraic tools to average over all possible tree-structured graphical models. The inferred network resulting from such an averaging procedure is not restricted to be a tree: species or groups of species can be isolated (e.g. Fig.~\ref{fig:networkinference}), and loops can appear (e.g. Fig.~\ref{fig:treeaveraging}). }

\modif{\paragraph{Network inference from species abundance data.}
This work focuses on network inference based on abundance data, and not only their presence/absence \citep[as considered in][]{OHS10,CWL18}.
Network inference from species abundance measures is a notoriously difficult problem \citep{ulrich2010null}, not only because network inference is complex, but also because it has to account for the data specificities. Abundance data may spread over a wide range of values and often result from sampling efforts (sample and/or species-specific), making them difficult to compare. 
\modif{Obviously, count data do not directly fit the Gaussian framework but many network inference methods dedicated to abundance data actually rely on a latent Gaussian structure (see Section \ref{altmethods}).}
%Any approach to inferring ecological networks needs to account for both sampling efforts and covariates describing the environment, as well as  distinguish between direct and indirect associations among species. \\
%\remove{Because the Gaussian setting is not suitable for count data, methods like gCoda \citep{gcoda} and SPIEC-EASI \citep{kurtz} resort to a continuous transformation of the data to then apply the glasso. In both methods, the sampling effort is accounted for by treating counts as compositional data, which does not model the randomness due to sampling in an explicit manner (proportions have a smaller variance when the sampling effort is greater).} 
}

\modif{\paragraph{Contribution.} 
In the present work, we adopt a model-based approach to perform network inference from abundance data. To accommodate the data specificities we use a PLN model, which includes the over-dispersion of the observed counts as well as the sampling effort. Importantly, the PLN model allows to account for abiotic effects and avoid the detection of spurious interactions between species.  \\
As for the network inference, we adopt a tree-based approach \citep[as opposed to][which also use a PLN model but resort to glasso]{MInt}, which provides a probability for each edge to be actually part of the underlying graphical model.}

\paragraph{Outline.}
We introduce the method EMtree, which combines two (variational) EM algorithms to estimate the model parameters. Importantly, our approach provides the probability for each possible edge to be part of the interaction network. We evaluate our approach on both synthetic and ecological datasets. An R package implementing EMtree is available on GitHub \url{https://github.com/Rmomal/EMtree}.  

%%%%%%%%%%%%%%%%%%%%%%%%%%%%%%%%%%%%%%%%%%%%%%%%%%%%%%%%%%%%%%%%%%%%%%%%

%Furthermore, indirect statistical association may be observed between two species and would interfere with interpretation. They can occur either because two species are both affected by the same environmental variations, or interacting with the same third species \citep{morueta2016network}. 
%\input{framework}

\section{Material and methods}
\subsection{Model} \label{sec:model} \modif{Let us first describe the typical type of data we consider.} We assume that $p$ species have been observed in $n$ sites and we denote $Y_{ij}$ the abundance of species $j$ in site $i$. The abundances are gathered in the $n \times p$ matrix $\Yb$. We denote by $\Yb_i$ the $i$th row of matrix $\Yb$, which corresponds to the abundance vector collected in site $i$. We further assume that a vector of covariates $\xb_i$ has been measured in each site $i$ and that all covariates are gathered in the $n \times d$ matrix $\Xb$. The sites are supposed to be independent.

Our aim is to decipher the dependency structure between the $p$ species, accounting for the effect of the environmental covariates encoded in $\Xb$. As explained above, ignoring environmental covariates is more than likely to result in spurious edges. 

\paragraph{Mixed model.}
To distinguish between covariates effects and species interactions, we consider a mixed model which states that each abundance $Y_{ij}$ has a (conditional) Poisson distribution
\begin{linenomath*}
\begin{equation} \label{eq:pY.Z}
    Y_{ij} \sim \Pcal\left(\exp(\xb_i^\intercal \thetab_j + o_{ij} + Z_{ij})\right).
\end{equation}
\end{linenomath*}
In model (\ref{eq:pY.Z}), $o_{ij}$ is the sample- and species-specific offset which accounts for the sampling effort. $\thetab_j$ is the vector of fixed regression coefficients measuring the effect of each covariate on species $j$ abundance. The regression part is similar to a general linear model as used in niche modeling \citep[see e.g.][]{austin2007species}. 
\modif{$Z_{ij}$ is the random effect associated with species $j$ in site $i$. 
Importantly, the coordinates of the site-specific random vector $\Zb_i = (Z_{i1}, \dots Z_{ip})$ are not independant: the multivariate random term $\Zb_i$ precisely accounts for the interactions that are not due to environmental fluctuations. 
For each site $i$, a vector $\Zb_i$ is associated with the corresponding abundance vector $\Yb_i$.}
%$Z_{ij}$ is a random effect, hence a random vector $\Zb_i = (Z_{i1}, \dots Z_{ip})$ (with {\sl dependent} coordinates) is associated with each abundance vector $\Yb_i$.  This random term precisely accounts for the species interactions that are not due to environmental fluctuations. 
%The dependency structure of the abundance vector $\Yb_i$ is then controlled by that of the latent vector $\Zb_i$. 
The distribution given in Eq.~\eqref{eq:pY.Z} is over-dispersed as the Poisson parameter is itself random, which suits ecological modeling of abundance data \citep{Eco_overdisp}.% as over-dispersion with respect to the Poisson distribution is often observed in abundance data \citep{Eco_overdisp}.
%It is worth noticing that the distribution \eqref{eq:pY.Z}  assumes that the counts $Y_{ij}$ are over-dispersed with respect to the regular Poisson distribution, as the Poisson parameter is itself random. 
% However, these dependency structures are not identical. 
% Even if the two dependency structures are not identical they share several common points of importance: the signs and null entries of their respective variance matrix are identical. Inferring the dependency structure  of the latent vector $\Zb_i$ is therefore highly informative about the dependence structure of the observed vector $\Yb_i$.

\bigskip
We now describe the distribution of the latent vector $\Zb_i$. To this aim, we adopt a version of Kirshner's model \citep{kirshner}, which states that a spanning tree $T$ is first drawn with probability
\begin{linenomath*}
\begin{equation} \label{eq:pT}
    p(T) = \prod_{(j, k) \in T} \beta_{jk} / B,
\end{equation}
\end{linenomath*}
where $(j, k) \in T$ means that the edge connecting species $j$ and $k$ is part of the tree $T$ and where $B$ is a normalizing constant. Each edge weight $\beta_{jk}$ controls the probability for the edge $(j, k)$ to be in the interaction network. \\
Then for each site $i$, a vector $\Zb_i$ is drawn independently with conditional Gaussian distribution $(\Zb_i \mid T) \sim \Ncal(0, \Sigma_T)$, where the subscript T means that the distribution of $\Zb_i$ is {faithful} to $T$. When $T$ is a spanning tree, this faithfulness simply means this distribution can be factorized on the nodes and edges of $T$ as follows \citep[see][]{kirshner}:
\begin{linenomath*}
\begin{equation} \label{eq:pZfact}
p(\Zb_i \mid T) = \prod_{j=1}^p p(Z_{ij}|T) \prod_{(j, k) \in T} \psi_{jk}(\Zb_i),
\end{equation}
\end{linenomath*}
where $\psi_{jk}(\Zb_i)$ does not depend on $T$. This factorization means that each edge of $T$ corresponds to a species pair in direct interaction;  all other pairs are conditionally independent. Experiments are independent, and in the sequel we consider the product of all $p(\Zb_i)$ and use the simpler notation $\psi_{jk} = \prod_i \psi_{jk}(\Zb_i)$ instead.\\

According to Eq.~\eqref{eq:pT}, each $\Zb_i$ has a Gaussian distribution conditional on the tree $T$, so its marginal distribution is a mixture of Gaussians: $\Zb_i \sim \sum_{T \in\mathcal{T}} p(T) \Ncal(0, \Sigma_T)$, where $\mathcal{T}$ is the set of all spanning trees. As a consequence, the joint distribution of the $\Zb_i$ is modeled by a mixture of distributions with tree-shaped dependency structure. \\
Besides, for all trees including the edge $(j, k)$, the estimate of the covariance term between the coordinates $j$ and $k$ is the same \citep[see][]{Lau96,SRS19}. Hence, we may define a global covariance matrix $\Sigmab$, filled with covariances that are each common to spanning trees containing a same edge. Each $\Sigmab_T$ is then built by extracting from $\Sigmab$ the covariances corresponding to the edges of $T$.

\subsection{Inference with EMtree} \label{sec:inference} We now describe how to infer the model parameters. We gather the edges weights $(\beta_\jk)_\jk$ into the $p \times p$ matrix $\betab$ and the vectors of regression coefficients into a $d \times p$ matrix $\thetab$. The $p \times p$ matrix $\Sigmab$ contains the variances and covariances between the coordinates of each latent vector $\Zb_i$. Hence, the set of parameters to be inferred is $(\betab, \Sigmab, \thetab)$.

\paragraph{Likelihood.} 
The model described above is an incomplete data model, as it involves two hidden layers: the random tree $T$ and the latent Gaussian vectors $\Zb_i$. The most classical approach to achieve maximum likelihood inference in this context is to use the Expectation-Maximization algorithm \citep[EM:][]{DLR77}. Rather than the likelihood of the observed data $p(\Yb)$, the EM algorithm deals with the often more tractable likelihood $p(T, \Zb, \Yb)$ of the complete data \modif{(which consists of both the observed and the latent variables)}. It can be decomposed as 
\begin{linenomath*}
\begin{equation} \label{eq:PTZY}
    p_{\betab, \Sigmab, \thetab}(T, \Zb, \Yb) = p_{\betab}(T) \times p_{\Sigmab}(\Zb \; | \; T) \times p_{\thetab}(\Yb \; | \; \Zb),
\end{equation}
\end{linenomath*}
where the subscripts indicate on which parameter each distribution depends. \\
Observe that the dependency structure between the species is only involved in the first two terms, whereas the third term only depends on the regression coefficients $\thetab$. 
We take advantage of this decomposition to propose a two-stage estimation algorithm. The first stage deals with the observed layer $p_{\thetab}(\Yb \; | \; \Zb)$, the second with the two hidden layers $p_{\betab}(T)$ and  $p_{\Sigmab}(\Zb \; | \; T)$. The network inference itself takes place in the second step.

\paragraph{Inference in the observed layer.} 
The variational EM (VEM) algorithm that provides an estimate of the regression coefficients matrix $\thetab$ is described in Appendix \ref{app:VEM} \modif{(along with a reminder on EM and VEM)}. It also provides the (approximate) conditional means $\Esp(Z_{ij} | \Yb_i)$, variances $\Var(Z_{ij} | \Yb_i)$ and covariances $\Cov(Z_{ij}, Z_{ik} | \Yb_i)$ required for the inference in the hidden layer. As a consequence, this first step provides the estimates $\widehat{\thetab}$ and $\widehat{\Sigmab}$.

%\modif{On dispose d'un algo VEM qui nous donne ($a$) une estimation de $\thetab$, et ($b$) les moments conditionnels (approchés) de $Z|Y$}

\paragraph{Inference in the hidden layer.} The second step is dedicated to the estimation of $\betab$. The EM algorithm actually deals with the conditional expectation of the complete log-likelihood, namely $\Esp\left(\log p_{\betab, \Sigmab, \thetab}(T, \Zb, \Yb) \; | \; \Yb\right)$. 
As shown in Appendix \ref{app:EM}, this  reduces to
\begin{linenomath*}
\begin{equation} \label{expectation}
    \Esp\left(\log p_{\betab, \Sigmab, \thetab}(T, \Zb, \Yb) \; | \; \Yb\right)
    \simeq
    \sum_{1 \leq j < k \leq p} P_\jk \log \left(\beta_\jk \widehat{\psi}_\jk\right) - \log B + \cst
\end{equation}
\end{linenomath*}
where $\widehat{\psi}_\jk$ is the estimate of $\psi_\jk$ defined in Eq.~\eqref{eq:pZfact}, and the '$\cst$' term depends on $\thetab$ and $\Sigmab$ but not on $\betab$. 
$P_\jk$ is the approximate conditional probability (given the data) for the edge $(j, k)$ to be part of the network:
$P_\jk \simeq \prob\{(j, k) \in T \; | \; Y\}$.
It is also shown in Appendix~\ref{app:EM} that $\widehat{\psi}_\jk = (1-\widehat{\rho}_\jk^2)^{-n/2}$, where the estimated correlation $\widehat{\rho}_\jk$ depends on the conditional mean, variance and covariances of the $Z_{ij}$'s provided by the first step.
 Eq.~\eqref{expectation} is maximized via an EM algorithm iterating the calculation of the $P_\jk$ and the maximization with respect to the $\beta_\jk$:
\begin{description}
\item[Expectation step: Computing the $P_\jk$ with tree averaging.] The conditional probability of an edge is simply the sum of the conditional probabilities of the trees that contain this edge. Hence, computing $P_\jk$ amounts to averaging over all spanning trees.
%The probability of an edge to be part of the network is the sum of the probabilities of all the trees containing this edge. 
Fig.~\ref{fig:treeaveraging} illustrates the principle of tree averaging for a toy network with $p=4$ nodes. Here, five arbitrary spanning trees $t_1$ to $t_5$ (among the $p^{p-2} = 16$ spanning trees) are displayed, with their respective conditional probability $p(T \mid Y)$. 
The edge $(1, 3)$ has a high conditional probability $P_{13}$ because it is part of likely trees such as $t_3$ and $t_4$, whereas $P_{23}$ is small because the edge $(2, 3)$ is only part of unlikely trees (e.g. $t_1$, $t_2$). \\
%As the number of possible trees is super-exponential in $p$, averaging  cannot be carried out in a naive way. However, thanks to the Matrix Tree theorem \citep[][recalled as Theorem~\ref{thm:MTT} in Appendix~\ref{app:MTT}]{matrixtree}  this is achievable at the cost of a determinant calculus (with complexity $O(p^3)$). 
%In practice, only combinations of the tree conditional probabilities $p(T \mid Y)$ are actually computed. 
\modif{Averaging over all spanning at the cost of a determinant calculus (i.e. with complexity $O(p^3)$) is possible using the Matrix Tree theorem \citep[][recalled as Theorem~\ref{thm:MTT} in Appendix~\ref{app:MTT}]{matrixtree}. }
\citet{kirshner} further shows that all the $P_\jk$'s can be computed at once with the same complexity $O(p^3)$, although the calculation may lead to numerical instabilities for large $n$ and $p$.

\begin{figure}[H]
    \centering
    \begin{tabular}{cccccc}
          \begin{tikzpicture}
  \node[observed] (1) at (0*\edgeunit, 0*\edgeunit) {$1$};
  \node[observed] (2) at (0*\edgeunit, 1*\edgeunit) {$2$};
  \node[observed] (3) at (1*\edgeunit, 1*\edgeunit) {$3$};
  \node[observed] (4) at (1*\edgeunit, 0*\edgeunit) {$4$};

  \draw[edge] (1) to (4); \draw[edge] (2) to (3); \draw[edge] (2) to (4);
  \end{tikzpicture} &
          \begin{tikzpicture}
  \node[observed] (1) at (0*\edgeunit, 0*\edgeunit) {$1$};
  \node[observed] (2) at (0*\edgeunit, 1*\edgeunit) {$2$};
  \node[observed] (3) at (1*\edgeunit, 1*\edgeunit) {$3$};
  \node[observed] (4) at (1*\edgeunit, 0*\edgeunit) {$4$};
  \draw[edge] (1) to (3); \draw[edge] (2) to (3); \draw[edge] (2) to (4);
  \end{tikzpicture} &
          \begin{tikzpicture}
  \node[observed] (1) at (0*\edgeunit, 0*\edgeunit) {$1$};
  \node[observed] (2) at (0*\edgeunit, 1*\edgeunit) {$2$};
  \node[observed] (3) at (1*\edgeunit, 1*\edgeunit) {$3$};
  \node[observed] (4) at (1*\edgeunit, 0*\edgeunit) {$4$};
  \draw[edge] (1) to (2); \draw[edge] (1) to (3); \draw[edge] (2) to (4);
  \end{tikzpicture} &
          \begin{tikzpicture}
  \node[observed] (1) at (0*\edgeunit, 0*\edgeunit) {$1$};
  \node[observed] (2) at (0*\edgeunit, 1*\edgeunit) {$2$};
  \node[observed] (3) at (1*\edgeunit, 1*\edgeunit) {$3$};
  \node[observed] (4) at (1*\edgeunit, 0*\edgeunit) {$4$};
  \draw[edge] (1) to (2); \draw[edge] (1) to (3); \draw[edge] (1) to (4);
  \end{tikzpicture} &
          \begin{tikzpicture}
  \node[observed] (1) at (0*\edgeunit, 0*\edgeunit) {$1$};
  \node[observed] (2) at (0*\edgeunit, 1*\edgeunit) {$2$};
  \node[observed] (3) at (1*\edgeunit, 1*\edgeunit) {$3$};
  \node[observed] (4) at (1*\edgeunit, 0*\edgeunit) {$4$};
  \draw[edge] (1) to (3); \draw[edge] (2) to (4); \draw[edge] (3) to (4);
  \end{tikzpicture} \\
%        $p(T_1 | Y) = 2.1\%$ & 
%        $p(T_2 | Y) = 3.5\%$ & 
%        $p(T_3 | Y) = 34.1\%$ & 
%        $p(T_4 | Y) = 15.6\%$ & 
%        $p(T_5 | Y) < .1\%$ \\ 
        $t_1: 2.1\%$ & 
        $t_2: 3.5\%$ & 
        $t_3: 34.1\%$ & 
        $t_4: 15.6\%$ & 
        $t_5:  <.1\%$ \\ 
        ~ \\
        & 
          \begin{tikzpicture}
  \node[observed] (1) at (0*\edgeunit, 0*\edgeunit) {$1$};
  \node[observed] (2) at (0*\edgeunit, 1*\edgeunit) {$2$};
  \node[observed] (3) at (1*\edgeunit, 1*\edgeunit) {$3$};
  \node[observed] (4) at (1*\edgeunit, 0*\edgeunit) {$4$};

  \draw[-, line width=2pt, color=black] (1) to (2); 
  \draw[-, line width=4pt, color=black] (1) to (3); 
  \draw[-, line width=2pt, color=black] (1) to (4);
  \draw[-, line width=.5pt, color=black] (2) to (3); 
  \draw[-, line width=4pt, color=black] (2) to (4); 
  \draw[-, line width=.1pt, color=black] (3) to (4);
  \end{tikzpicture}

% > round(50*Pedge)/10
%      [,1] [,2] [,3] [,4]
% [1,]  0.0  2.6  4.8  3.0
% [2,]  2.6  0.0  0.5  4.1
% [3,]  4.8  0.5  0.0  0.0
% [4,]  3.0  4.1  0.0  0.0
% > Pedge
%           [,1]       [,2]         [,3]         [,4]
% [1,] 0.0000000 0.52246119 9.540435e-01 6.057683e-01
% [2,] 0.5224612 0.00000000 9.751246e-02 8.201690e-01
% [3,] 0.9540435 0.09751246 0.000000e+00 4.565667e-05
% [4,] 0.6057683 0.82016896 4.565667e-05 0.000000e+00
 &
        \qquad \qquad &
          \begin{tikzpicture}
  \node[observed] (1) at (0*\edgeunit, 0*\edgeunit) {$1$};
  \node[observed] (2) at (0*\edgeunit, 1*\edgeunit) {$2$};
  \node[observed] (3) at (1*\edgeunit, 1*\edgeunit) {$3$};
  \node[observed] (4) at (1*\edgeunit, 0*\edgeunit) {$4$};

  \draw[edge] (1) to (2); \draw[edge] (1) to (3); \draw[edge] (1) to (4); \draw[edge] (2) to (4);
  \end{tikzpicture} \\
        \multicolumn{3}{c}{Edge conditional probabilities} & Estimated graph \\
    \end{tabular}
    \caption{Tree averaging principle. 
    \textit{Top:} \modif{a subset of 5 spanning trees with 4 nodes  $(t_1, \dots t_5)$, with their respective conditional probability given the data $P(T = t \mid Y)$.} 
    \textit{Bottom left:} The weighted graph resulting from tree averaging. Each edge $(j, k)$ has width proportional to its conditional probability. \textit{ Bottom right:} The estimated graph (obtained by thresholding edge probabilities) is not a tree.}
    \label{fig:treeaveraging}
\end{figure}
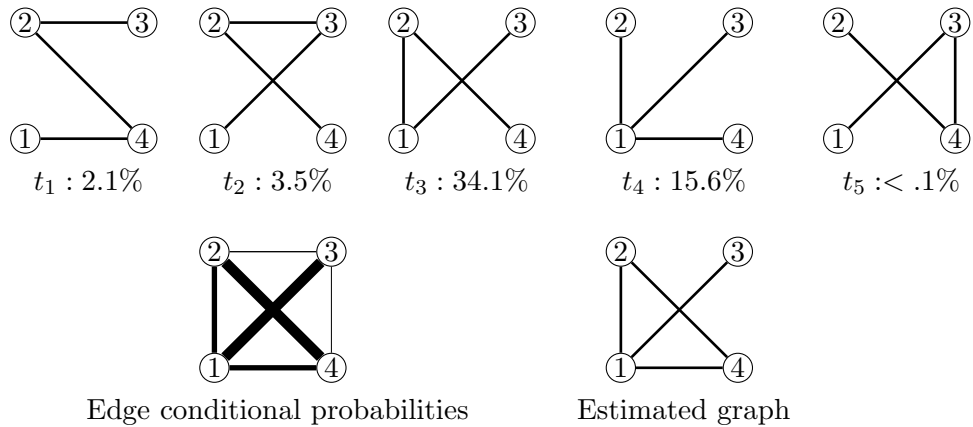

\item[Maximization step: Estimating the $\beta_\jk$.] 
\modif{This step is not straightforward, as the normalizing constant $B = \sum_T \prod_{(j, k) \in T} \beta_\jk$ involves all $\beta_\jk$'s. We propose an exact maximization built upon the Matrix Tree theorem (see Appendix~\ref{app:EM}). }
%Maximizing Eq.~\eqref{expectation} with respect to each of the $\beta_\jk$ is not straightforward, as the normalizing constant $B = \sum_T \prod_{(j, k) \in T} \beta_\jk$ involves all $\beta_\jk$'s. Still, the gradient of Eq.~\eqref{expectation} can be computed using Lemma 2 from \citet{MeilaJaak} (recalled as Lemma~\ref{lem:Meila} in Appendix~\ref{app:MTT}), which also builds upon the Matrix Tree theorem (see Appendix~\ref{app:EM}).
\end{description}

\paragraph{Algorithm output: edge scoring and network inference} 
%As a side product, 
EMtree provides the (approximate) conditional probability $P_\jk$ for each edge $(j, k)$ to be part of the network. 
%Hence, an estimate $\widehat{G}$ of the network can be easily obtained by thresholding conditional probabilities: for a given threshold $0 < \lambda < 1$, the edge $(j, k)$ will be in $\widehat{G}$ if $P_\jk \geq \lambda$. Fig.~\ref{fig:treeaveraging} (bottom right) illustrates that the resulting graph has no reason to be a tree. As our model is build on a tree hypothesis, a natural rule is to select the edges with probability $P_\jk$ greater than the probability for an edge to be part of a tree drawn uniformly, that is $\lambda = 2/p$.
\modif{Whenever an actual inferred network $\widehat{G}$ is needed (e.g. for a graphical purpose), it can be obtained by thresholding the $P_\jk$ (see Fig.~\ref{fig:treeaveraging}, bottom right). Because we are dealing with trees, a natural threshold is the density of a spanning tree, which is  $2/p$.}
%To get a more robust estimate $\widehat{G}$, a resampling procedure can be designed in the spirit of the stability selection proposed by  \citet{LRW10}. $S$ sub-samples are drawn, each using a fraction $f$ of available observations. For each sub-sample $s = 1 \dots S$, an estimate $\widehat{G}^s$ is made of the edges having probability $P^s_\jk \geq  2/p$. Only edges selected in more than a fraction $f'$ of the estimated graphs $\widehat{G}^s$ are kept to build the final $\widehat{G}$. This procedure can be parallelized in an obvious manner. In all simulations and applications, we set $f = 80\%$. $f'$ is a more flexible threshold; we set $f'=80\%$ in all simulations, and $f'=90\%$ in all applications.
\modif{More robust results can be obtained using a resampling procedure similar to the stability selection proposed by \citet{LRW10}. It simply consists in sampling a series of subsample $s = 1 \dots S$, to get an estimate $\widehat{G}^s$ from each of them and to collect the selection frequency for each edge. Again, these edge selection frequencies can be thresholded if needed.}
\subsection{Simulation and illustrations} 
\modif{Because network inference is an unsupervised problem (as opposed to network reconstruction), we compare the accuracy of the methods described above on synthetic abundance datasets, for which the true underlying network is known.}

\subsubsection{Alternative inference methods} \label{altmethods}

\modif{
We consider network inference methods dedicated to both 
metagenomics (SPIEC-EASI, gCoda and MInt) and ecology (MRFcov, ecoCopula). All methods can handle count data and rely on some (implicit) Gaussian setting. SPIEC-EASI \citep{kurtz}, gCoda \citep{gcoda} and MRFcov \citep{CWL18} resort to data transformation to fit a Gaussian framework. MInt \citep{MInt} considers a Poisson mixed model similar to the one  we consider and ecoCopula \citep{PWT19} defines a multivariate count distribution, the dependency structure of which is encoded in a Gaussian copula. These methods all rely on a Gaussian graphical model (GGM) or a Gaussian copula, so that the network inference problem amounts to estimating a sparse version of the inverse covariance matrix (also named {\sl precision} matrix). 
 }

\paragraph{Edge scoring.}
 These methods build upon glasso penalization \citep{FHT08}. For each edge, there exists a minimal penalty value above which it is eliminated from the network. The higher this minimal penalty, the more reliable the edge in the network, so it can be used as a score reflecting the importance of an edge. \modif{Only SpiecEasi and gCoda provide unthresholded quantities (namely the glasso regularization path) that can be used for edge scoring; the other methods only return their optimal graph.}
 \modif{\paragraph{Covariates.} Only MInt, MRFcov and ecoCopula may include covariates. In order to draw a fair comparison, we give SPIEC-EASI and gCoda access to the covariate information by feeding them with residuals of the linear regression of the transformed data onto the covariates.}

\subsubsection{Comparison criteria}

\paragraph{False Discovery Rate (FDR) and density ratio criteria.}
Inferred networks are mostly useful to detect potential interactions between species, which then need to be \modif{studied by experts} to determine their exact nature. Falsely including an edge lead to \modif{meaningless interpretation} \modif{or useless validation experiments}. 
A network with a few reliable edges will be preferred to one having more edges with a larger risk of possible false discoveries. Therefore we choose the FDR as an evaluation criterion, which should be close to 0. Comparing FDR's only makes sense for networks with similar densities. We then compute the ratio between the densities of the inferred and the true network ({\sl density ratio}).

\paragraph{Area Under the Curve (AUC) criterion.}
The AUC criterion allows to evaluate the inferences quality
\modif{without resorting to any threshold}. It evaluates the probability for a method to score the presence of a present edge higher than that of an
absent one; it should be close to 1. Note that this criterion \modif{ cannot be computed for MRFcov, ecoCopula and MInt as they provide} a unique 
\modif{network}.

%%%%%%%%%%%%%%%%%%%%%%%%%%%%%%%%%%%%%%%%%%%%%%%%%%%%%%%%%%
%%%%%%%%%%%%%%%%%%%%%%%%%%%%%%%%%%%%%%%%%%%%%%%%%%%%%%%%%%

\subsubsection{Simulation design}
 
\paragraph{Simulated \modif{graphs}.}
We consider three typical graph structures: Scale-free, Erdös (short for Erdös-Reyni) and Cluster.
Scale-free structure bears the closest similarity to the tree one, with almost the same density and no loops; it is popular in social networks and in genomics as it corresponds to a preferential-attachment behavior. 
It is simulated following the Barb\'{a}si-Albert model as implemented in the \textit{huge} R package \citep{huge}. The degree distribution of Scale-free structure follows a power law, which constrains the edges probabilities such that the network density cannot be controlled.
Erdös structure is the most even as the edges all have the same existence probability. It is a step away from the tree as it may contain loops and its density can be increased arbitrarily.
Cluster structure spreads edges into highly connected clusters, with few connections between the clusters; the \textit{ratio} parameter controls the intra/inter connection probability ratio.

\paragraph{\modif{Simulated counts.}}
\modif{The datasets are simulated under the Poisson mixed model described in Eq.~\eqref{eq:pY.Z}. We first build the covariance matrix $\Sigma_G$ associated with a graph $G$ following \citet{huge} and randomly choosing the sign of the link, so that in our simulations we consider both positive and negative interactions. For each site $i$, we simulate $\Zb_i \sim \Ncal(0,\Sigma_G)$, then use these parameters together with a set of covariates to generate count data $\Yb$. We use three covariates (one continuous, one ordinal and one categorical) to create a heterogeneous environment.}

\paragraph{\modif{Experiments}.}
For each set of parameters and type of structure we generate 100 graphs, simulate a dataset under a heterogeneous environment and infer the dependency structure using EMtree, gCoda, SpiecEasi \modif{MInt, ecoCopula and MRFcov (the three latter only for Exp. 1)}. 
\modif{The settings of all methods are set to default, except for ecoCopula for which we use the "AIC" selection criterion ("BIC" gives too many empty results).}
All computation times are obtained with a 2.5 GH Intel Core 17 processor and 8G of RAM.
\modif{
\begin{description}
\item[Exp. 1: effect of the data dimensions on the inferred network.] We compare performances in terms of FDR and density ratios on two scenarios: \textit{easy} ($n=100$, $p=20$), and \textit{hard} ($n=50$, $p=30$). The network density for Erdös and Cluster structures is set to $\log(p)/p$.
\item[Exp. 2: effect of the network structure on edge rankings.] AUC measures are collected for alternate variations of $n$ and $p$ to get a general idea of each performance. For comparison's sake, the same density is fixed for all structures in this case,  so that only $n$ and $p$ vary in turn; the scale-free structure imposes a common density of $2/p$. The default values are $n=100$, $p=20$. 
\item[Exp. 3:  effect of the graph density on edge rankings.] AUC measures are collected for variations of $n$ and $p$ with a density of $5/p$ (5 neighbors per node on average), and for variations of density parameters. The default values are $n=100$, $p=20$.
\end{description}
}

 \subsubsection{Illustrations} \label{sec:datasets}
 
The first application deals with fish population measurements in the estuary of the Fatala River, Guinea, \citep[][available in the R package \textit{ade4}]{baran1995dynamique}. The data consists of 95 count samples of 33 fish species, and two covariates {\it date} and {\it site}. 
We infer the network using four models including no covariates, either one  or both covariates (i.e. respectively the \textit{null}, \textit{site}, \textit{date} and \textit{site+date} models)

The second example is a metabarcoding experiment designed to study oak powdery mildew \citep{jakuch}, caused by the fungal pathogen \textit{Erysiphe alphitoides} (Ea). To study the pathobiome of oak leaves, measurements were done on three trees with different infection status. The resulting dataset is composed of 116 count samples of 114 fungal and bacterial  operational taxonomic units (OTUs) of oak leaves, including the Ea agent.  \modif{The original raw data are available at \url{https://www.ebi.ac.uk/ena/data/view/PRJEB7319}}. Several covariates are available, among which the tree status, the orientation of the branch, and three covariates measuring the distances of oak leaves to the ground (D1), to the base of the branch (D2), and to the tree trunk (D3).\modif{ The experiment used different depths of coverage for bacteria and fungi, which we account for via the offset term. We fitted three Poisson mixed models including either none, the tree status or all of the covariates (i.e. respectively \textit{null}, \textit{tree}, and \textit{tree+D1+D2+D3} models).}

To further analyze the inferred networks, we use the betweenness  centrality \citep{centrality}, a centrality measure popular in social network analysis. It measures a node's ability to act as a bridge in the network. High betweenness scores  identify sensitive nodes that can efficiently describe a network structure. We compute these using the R package \textit{igraph}.

\section{Results}
\subsection{On simulated data} \subsubsection{Effect of dataset dimensions}
\label{adverse}

\begin{figure}[H]
    \centering
    \includegraphics[width=\linewidth]{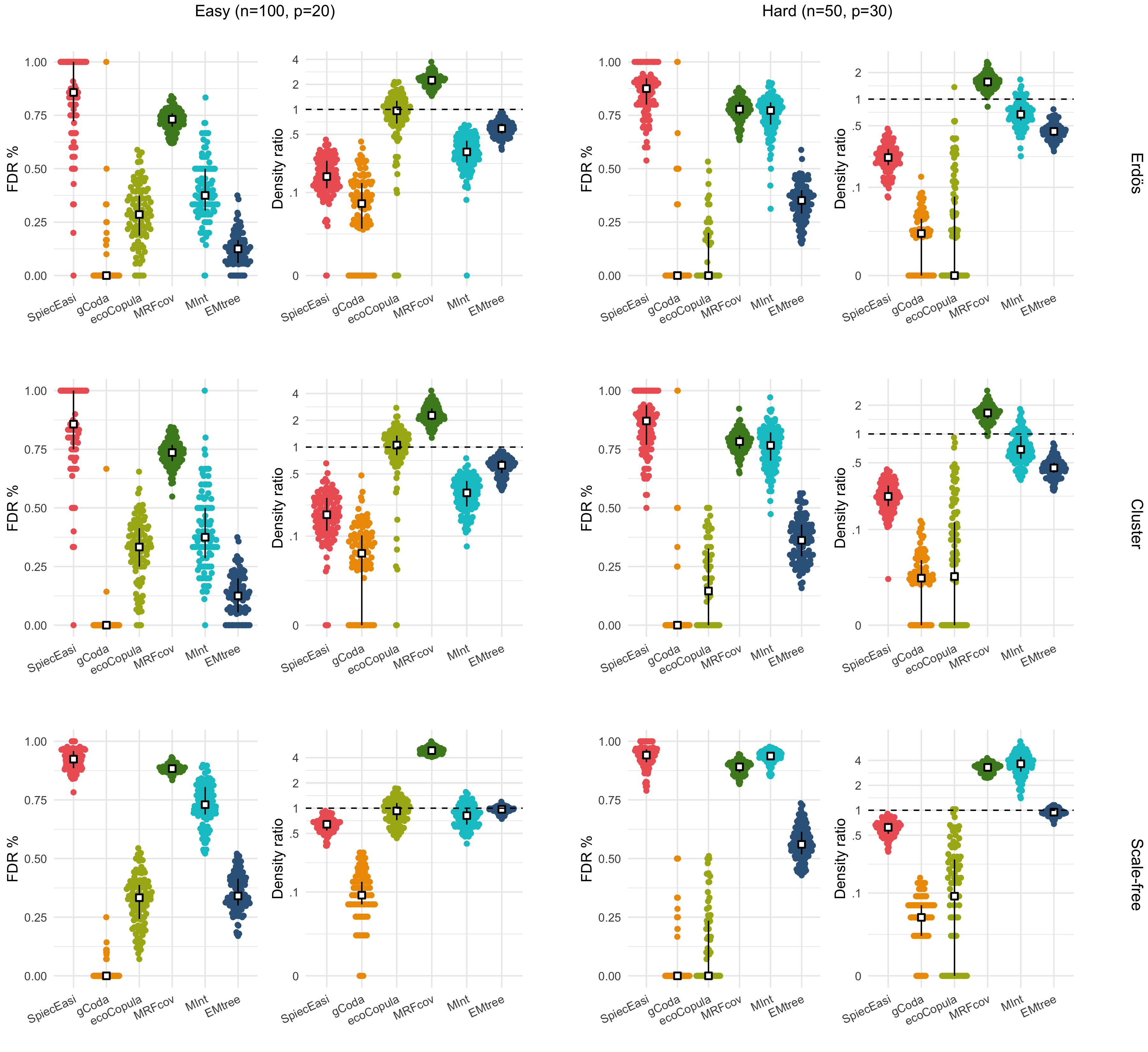}
    \caption{FDR and density ratio measures for all methods at two different difficulty levels and 100 networks of each type. White squares and black plain lines represent medians and quartiles respectively. \small{\textit{ecoCopula selection method: AIC. Number of subsamples for SpiecEasi and EMtree: $S=20$. SpiecEasi and gCoda: $lambda.min.ratio=0.001$,  $nlambda=100$.}}}
    \label{TPFN}
\end{figure}

Behaviors are compared on an easy setting ($n=100$, $p=20$) and a hard setting ($n=50$, $p=30$). Fig.~\ref{TPFN} displays FDR and density ratio measures for all methods on the different cases.  Detailed values of medians and standard-deviations are given in Tables \ref{medFDR} and\ref{meddens}. \modif{The behaviour of methods remains virtually the same across Erdös and Cluster structures. Scale-free structure appears to entail a greater difficulty for all methods except ecoCopula: the FDR increases in easy cases of about $15\%$ for SpeacEasi, MRFcov and EMtree, and about $35\%$ for MInt.\\
The greater difficulty affects all methods. gCoda standard-deviation increases by $10\%$. MRFcov, EMtree and MInt show an increase in FDR of about $5\%$, $20\%$ and $30\%$ respectively. Density ratios overall decrease, especially for ecoCopula which ratio is close to 0 and yields a proportion of empty networks of $15$-$25\%$ (Table \ref{empty})}.

\modif{Considering FDRs and density ratios combined, EMtree appears to be the method with the lower FDR which maintains a density ratio reasonably close to 1. As a consequence, the proposed methodology compares well to existing tools on problems with varying difficulties. EMtree is also comparable on running times. Table~\ref{timesTPFN} shows that for Erdös and Cluster it is the third quicker method in easy cases and the second in hard ones. Table \ref{timeSF} (in appendix) shows that on scale-free problems, EMtree is the second quicker method in hard cases, and is curiously slow on easy ones.}

Interestingly, in easy cases when the network density is well estimated, methods yield FDR of $10\%-30\%$ in median. This reminds that network inference from abundance data is a difficult task, and that perfect inference of the network remains an out-of-reach goal.

% Fig.~\ref{Seffect} shows EMtree FDR and density ratio measures for $S$ between 1 and 150. Clearly, the performances stabilize from about  $S=10$. Therefore the same results can be reached with much lower $S$ values. For example $S=20$ seems a reasonable choice of settings here, and running times would then be lower than those of MInt in all cases according to Table~\ref{timesS}.

\begin{table}[ht]
\centering
%\begin{tabular}{l|rlrlrlrlrlrl}
% & \multicolumn{2}{c}{SpiecEasi} & \multicolumn{2}{c}{gCoda} & \multicolumn{2}{c}{ecoCopula} & \multicolumn{2}{c}{MRFcov} & \multicolumn{2}{c}{MInt} & \multicolumn{2}{c}{EMtree} \\ 
%  \hline
%Easy & 25.45 & (1.87) & 0.11 & (0.06) & 5.55 & (0.64) & 34.51 & (3.68) & 43.04 & (19.76) & 11.72 & (1.89) \\ 
%  Hard & 28.43 & (1.30) & 0.53 & (0.25) & 9.6 & (0.65) & 8.29 & (0.36) & 33.77 & (18.20) & 8.17 & (0.50) \\ 
%   \hline
%\end{tabular}
\begin{tabular}{l|rrrrrr}
 & \multicolumn{1}{c}{SpiecEasi} & \multicolumn{1}{c}{gCoda} & \multicolumn{1}{c}{ecoCopula} & \multicolumn{1}{c}{MRFcov} & \multicolumn{1}{c}{MInt} & \multicolumn{1}{c}{EMtree} \\ 
  \hline
Easy & 25.45  (1.87) & 0.11  (0.06) & 5.55  (0.64) & 34.51  (3.68) & 43.04  (19.76) & 11.72  (1.89) \\ 
  Hard & 28.43  (1.30) & 0.53  (0.25) & 9.6  (0.65) & 8.29  (0.36) & 33.77  (18.20) & 8.17  (0.50) \\ 
   \hline
\end{tabular}
\caption{Median and standard-deviation running-time values (in seconds) for Cluster and Erdös structures, including resampling with $S=20$ for SpiecEasi and EMtree.}
\label{timesTPFN}
\end{table}

%%%%%%%%%%%%%%%%%%%%%%%%%%%%%%%%%%%%
%%%%%%%%%%%%%%%%%%%%%%%%%%%%%%%%%%%%

\subsubsection{Effect of network structure}

\begin{figure}[H]
    \centering
    
    \includegraphics[width=0.7\linewidth]{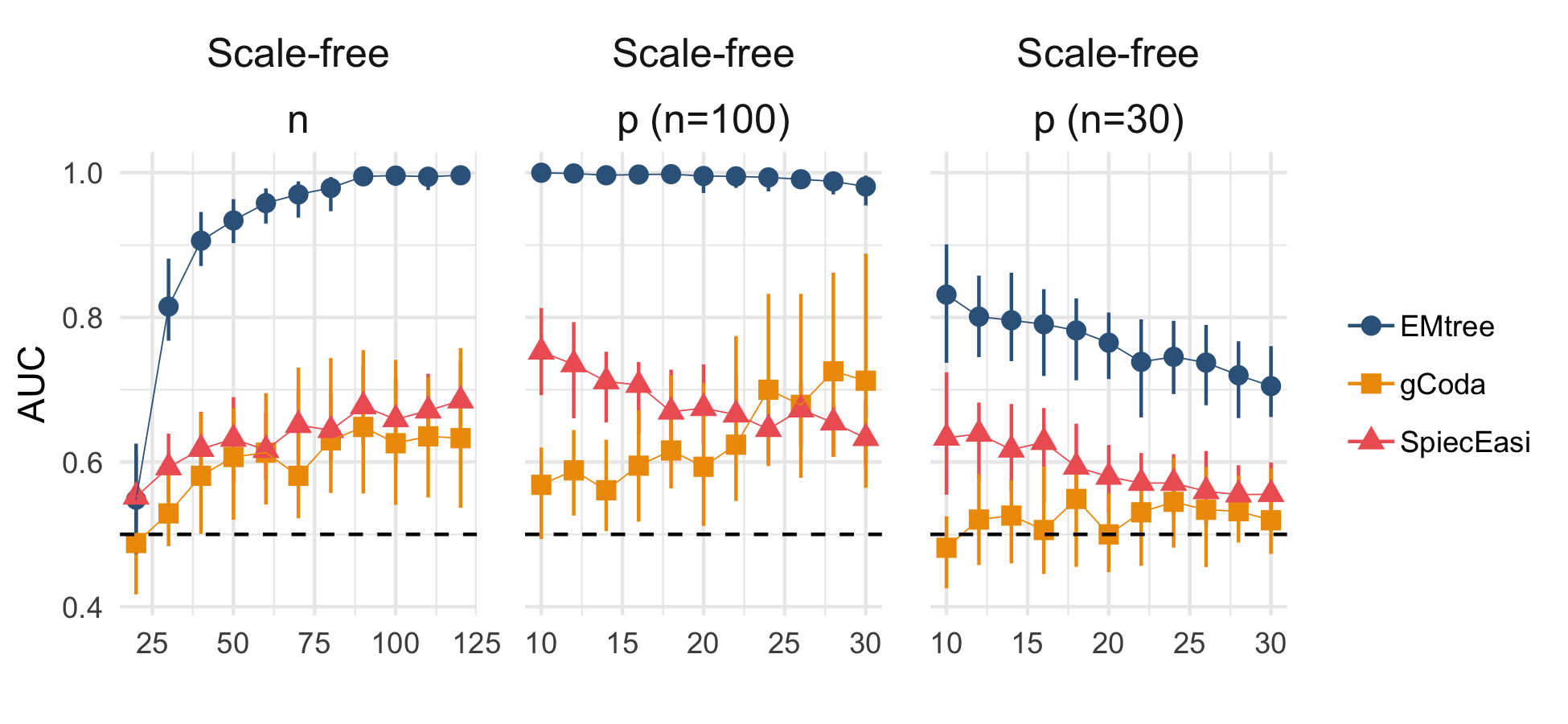}
    \caption{Effect of Scale-free structure on AUC medians and inter-quartile intervals for parameters $n$ and $p$.}
      \label{SFAUC}
\end{figure}

As expected for a fixed $p$, the higher the number of observations $n$, the better the performance for all methods and structures. Interestingly, the same happens when $p$ increases for a fixed $n=100$ (except for SpiecEasi).
EMtree performs well on Scale-free structures (Fig.~\ref{SFAUC}) which was also expected; the other methods performance worsen compared to other structures. When lowering $n$ to 30, EMtree performance deteriorates along with $p$, yet remaining above $70\%$ in median in the extreme case where $p=n$ (Fig.~\ref{SFAUC}, right). The structure being Erdös or Cluster, each method is affected in the same way by an increase of $n$ or $p$ (Fig.~\ref{panelErdClust}). Besides, increasing the difference between the two structures by tuning up the \textit{ratio} parameter has no effect. Overall EMtree performs better than gCoda and SpiecEasi on all the studied configurations. Running times are summarized in Table~\ref{timeNP}. EMtree is about 10 times slower than gCoda (4 for small $n$), and 4 times faster than SpiecEasi. The high standard deviation for small $n$ seems to be due to gCoda struggling with Scale-free structures.
 
%\modif{ EMtree shows almost perfect performance on tree-like structures. Nevertheless, this method overall performs better than gCoda and SpiecEasi on all the structures studied.  }

\begin{table}[H]
\centering
\begin{tabular}{l|rr|rr}
 & \multicolumn{1}{c}{$n < 50$} & \multicolumn{1}{c}{$n\geq 50$}  & \multicolumn{1}{c}{$p < 20$} & \multicolumn{1}{c}{$p\geq 20$} \\  
 \hline
  EMtree    &   0.44 (0.14)	 &   0.60 (0.17) &   0.41 (0.13) &   0.76 (0.21)   \\ 
  gCoda     &   0.11 (26.8)	 &   0.05 (0.05) &   0.05 (0.04) &   0.09 (0.54)   \\ 
  SpiecEasi &   2.09 (0.26)	 &   2.37 (0.28) &   2.42 (0.27) &   2.42 (0.26)   \\ 
   \hline
\end{tabular}
\caption{Median and standard-deviation of running times for each method in seconds, for $n$ and $p$ parameters.}
\label{timeNP}
\end{table}

%%%%%%%%%%%%%%%%%%%%%%%%%%%%%%%%%%%%
%%%%%%%%%%%%%%%%%%%%%%%%%%%%%%%%%%%%

\subsubsection{Effect of network density}
\begin{figure}[H]
  \centering
   \includegraphics[width=\linewidth]{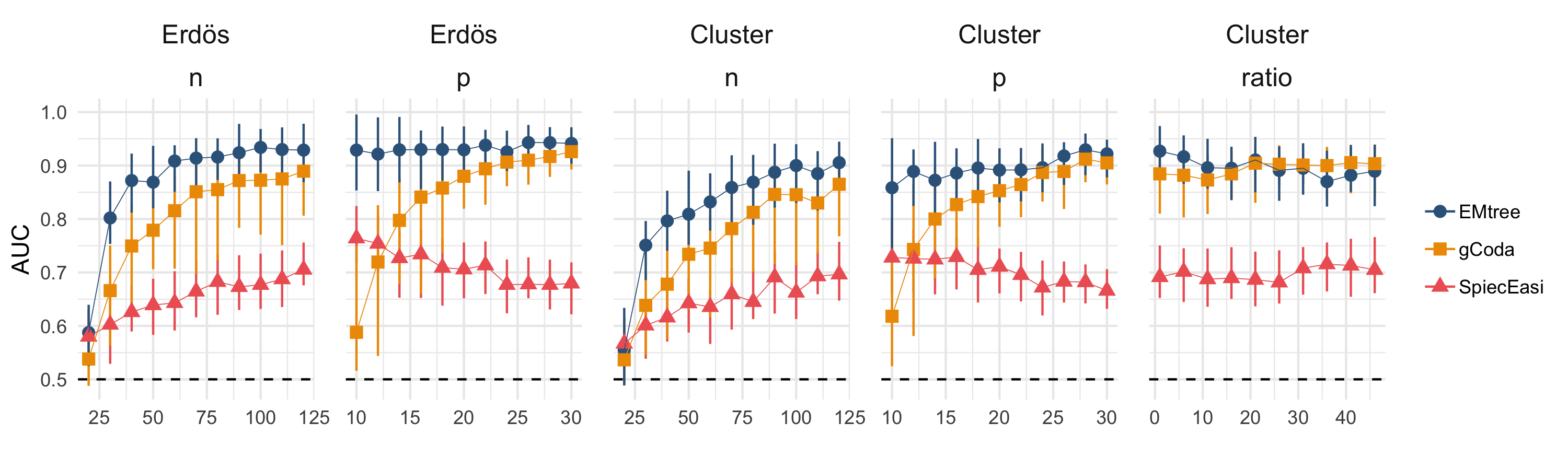}
  \includegraphics[width=\linewidth]{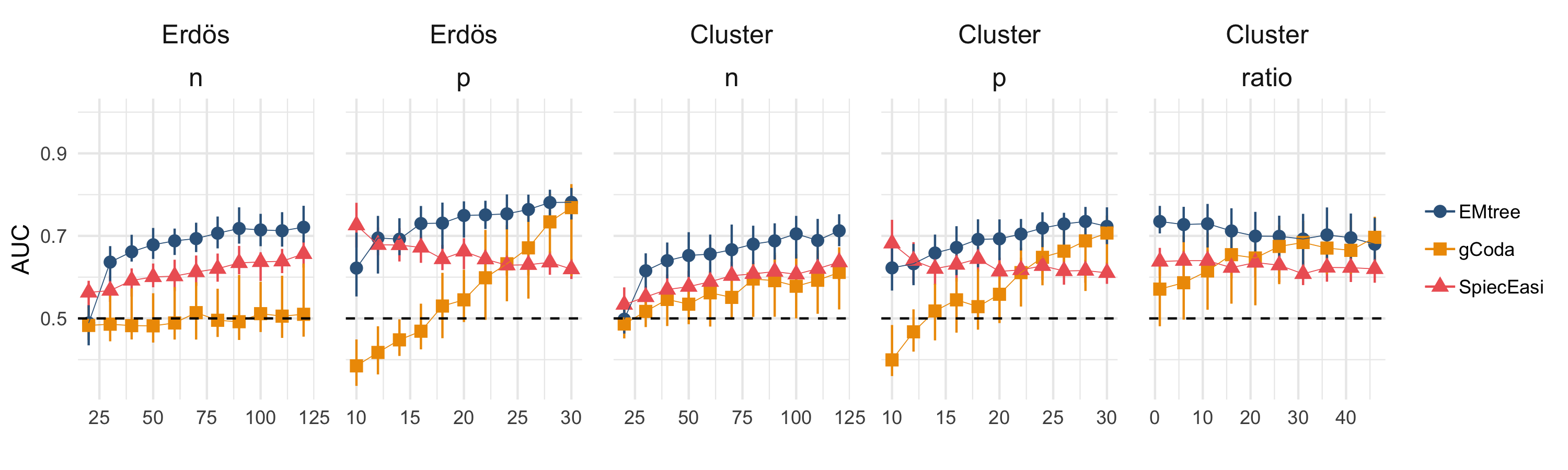}
  \caption{Effect of Erdös and Cluster structures on AUC medians and inter-quartile intervals for parameters $n$, $p$ and $ratio$. \textit{Top}: densities set to $2/p$, \textit{bottom}: densities set to $5/p$.}
  \label{panelErdClust}
\end{figure}

\begin{figure}[H]
 \centering
  \includegraphics[width=0.6\linewidth]{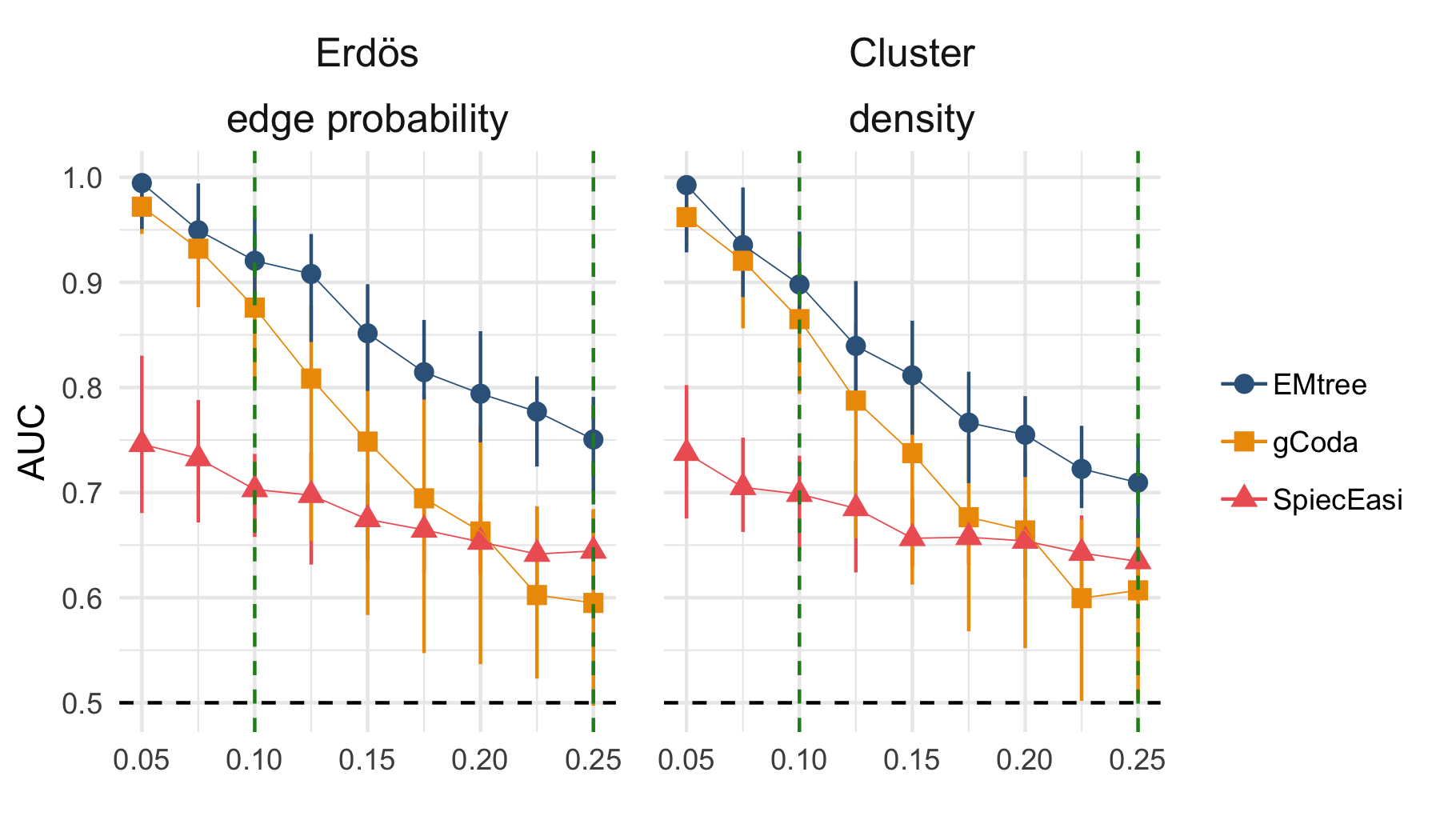}
  \caption{AUC median and inter-quartile intervals for parameters controlling the number of edges in Erdös (\textit{edge probability}) and in Cluster (\textit{density}) structures, $p=20$, $n=100$.}
  \label{varyDens}
\end{figure}

The comparison of top and bottom panels of Fig.~\ref{panelErdClust} shows that network inference gets harder as the network gets denser, whatever the method and the structure of the true graph. Running times are not affected (Table \ref{timeDenser}).
Fig.~\ref{varyDens} shows that EMtree performance does not deteriorate faster than that of other methods, demonstrating that the tree hypothesis is not a limitation.

 %Fig.~\ref{varyDens} summarizes the AUC measures for a varying graph density, with fixed parameters $n$ and $p$. It shows in details the performance degradation for all methods along with the network density.  Fig.~\ref{panelErdClust} (bottom) presents AUC measures for variations of $n$ and $p$ parameters with an increased density of $5/p$.  The graph density does not seem to impact running times for parameters $n$ and $p$ (not shown). This highlights that  \\}
 %gCoda performance deteriorates faster than that of the other methods: its outcomes are close to EMtree's for a $2/p$ density, and below SpiecEasi's for a $5/p$ density. 
%\modif{
%Overall when the density varies EMtree still compares well to the other methods which do not have an underlying tree structure hypothesis. This demonstrates that the tree hypothesis of our model is not a limitation.}

 \label{sec:simul}
\subsection{Illustrations} In this section we emphasize the importance of covariates for network inference. Accounting for environmental effects changes the structure of all inferred networks we present; nodes with the highest betweenness scores are highlighted to spot these changes. Most frequently, it results in reducing the number of edges (i.e. making the network sparser). However new edges can appear as well, as adjusting for a covariate also reduces the variability, which improves the detection power. In all examples, we used the resampling method described in Section \ref{sec:inference}, which provides edge selection frequencies. Eventually, we have to threshold these frequencies to draw actual networks; the value of the threshold obviously affects the density of the plotted networks (see Fig.~\ref{QETOak}). 

%\remove{The curves on this figure are very smooth, illustrating the difficulty of setting this threshold.}

\subsubsection{Fish populations in the Fatala River estuary}
\label{barans}

\begin{figure}[H]
      \includegraphics[width=\linewidth]{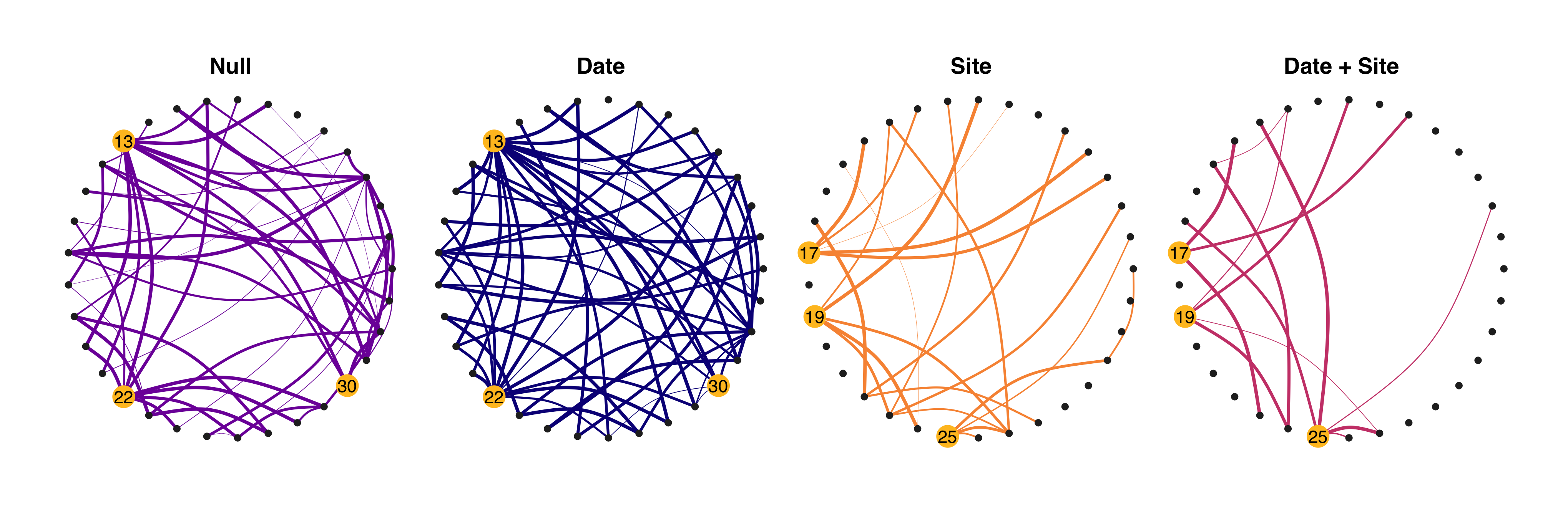}
      \caption{Interaction networks of Fatala River fishes inferred when adjusting for none, both or either one of the covariates among \textit{site} and \textit{date}. Highlighted nodes spot the highest betweenness centrality scores. Widths are proportional to selection frequencies. $S=100$, $f'=90\%$. }
    \label{baransNets}
  \end{figure}

%The networks shown in Fig.~\ref{baransNets} were obtained with $S=100$ sub-samples, each of which took about $0.3s$ to be inferred with EMtree. On each network, the three species with the highest betweenness scores are highlighted; the corresponding species names are given in Appendix~\ref{names_Baran}. \\
Networks on Fig.~\ref{baransNets} suggest a predominant role of the \textit{site} covariate compared to the \textit{date}. Indeed, adjusting for the \textit{site} results in much sparser networks (Fig.~\ref{QETOak} in appendix). It deeply modifies the network structure: the \textit{site} network has 12 new edges and only 6 in common with the \textit{null} network. Besides, the  highlighted nodes only change when introducing the \textit{site} covariate.  This suggests that the environmental heterogeneity between the sites has a major effect on the \modif{variations of species abundances}, while the effect of the date of sampling is moderate.

%%%%%%%%%%%%%%%%%%%%%%%%%%%%%%%%%
\subsubsection{Oak powdery mildew}  
\label{oak}

%The networks presented on Fig.~\ref{oakNets} were obtained using $S=100$  sub-samples as well. Running EMtree on each of them took about 14s. Each network highlights the six nodes with the highest betweenness scores and the connections of Ea.

When providing the inference with more information (tree status, distances), the structure of the resulting network is significantly modified. Nodes with high betweenness scores differ from one model to another. There is an important gap in density between the \textit{null} model and the others, starting from a $25\%$ selection threshold (Fig.~\ref{QETOak} in appendix). From a more biological point of view, the features of the pathogen node are greatly modified too: its betweenness score is among the smallest in the \textit{null} network (quantile $16\%$), and among the highest in the two other networks (quantiles $93\%$ and $96\%$). Its connections to the other nodes vary as well. 
%So here taking covariates into account results in less interactions with the pathogen, and a greater role of the latter in the pathobiome organization.
\modif{
Accounting for covariates results in less interactions with the pathogen but a greater role of the latter in the pathobiome organization.
}

 \begin{figure}[H]
    \centering
    \includegraphics[width=\linewidth]{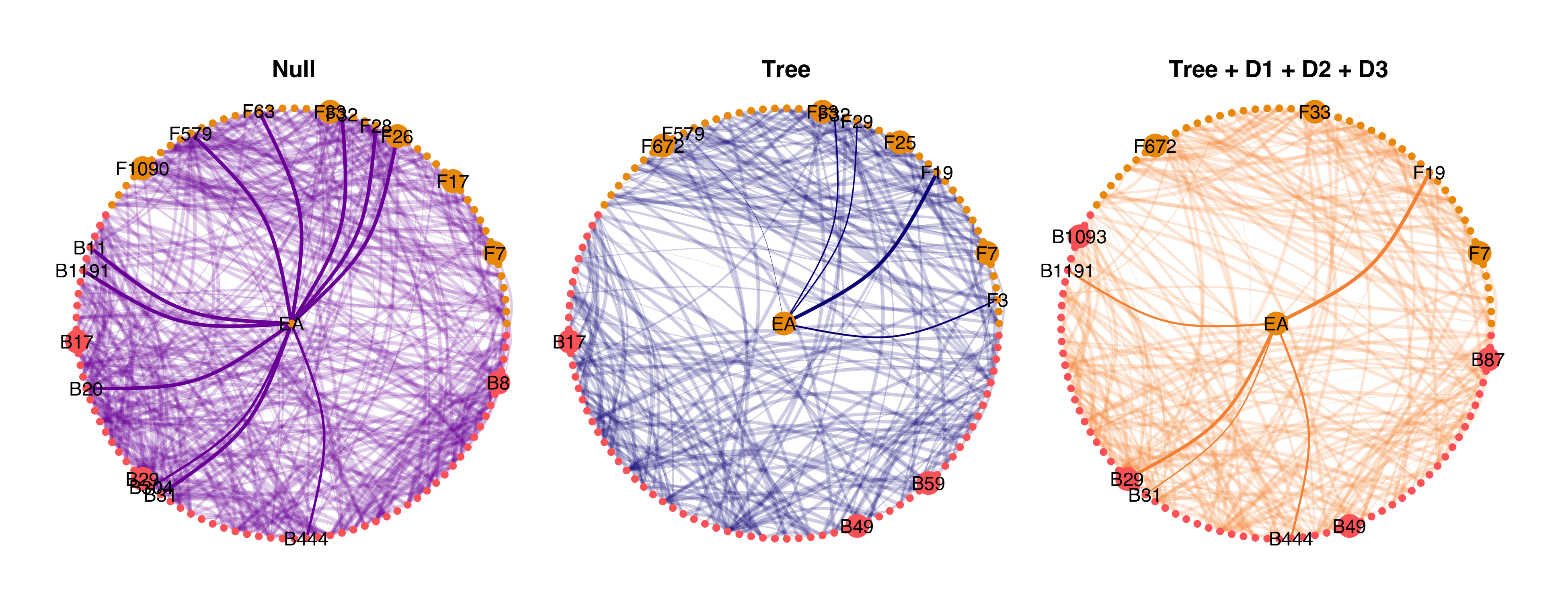}
    \caption{Pathogen interaction networks on oak leaves inferred with EMtree when adjusting for none, the \textit{tree} covariate or \textit{tree} and distances. Bigger nodes represent OTUs with highest betweenness values, colors differentiate fungal and bacterial OTUs. Widths are proportional to selection frequencies. $S=100$, $f'=90\%$ .  }
    \label{oakNets}
\end{figure}

%[ ] Comparison avec les résultats de jakuschkin :
  %  - [ ] Présentation de jakuschkin 
  %  - [ ] On compare nos résultats aux leurs en faisant varier le seuil de %sélection
%Following a Bayesian approach, \citet{jakuch} identifies a list of 26 OTUs likely to be directly interacting with Ea, as well as a list of 34 OTUs significantly associated with infected samples. We compare this list of 52 OTUs to EMtree results when varying the selection threshold $f'$. On Fig.~\ref{compareOTUs} the fraction of OTUs identified by EMtree and  shared with the previous study seems to stabilize around $70\%$, showing these outcomes reasonable consistency. Note that for EMtree to identify about the same quantity of OTUs as \citet{jakuch}, the selection threshold must be set between $10\%$ and $15\%$, which seems low to us.

\begin{figure}[H]
    \centering
    \includegraphics[width=9.5cm]{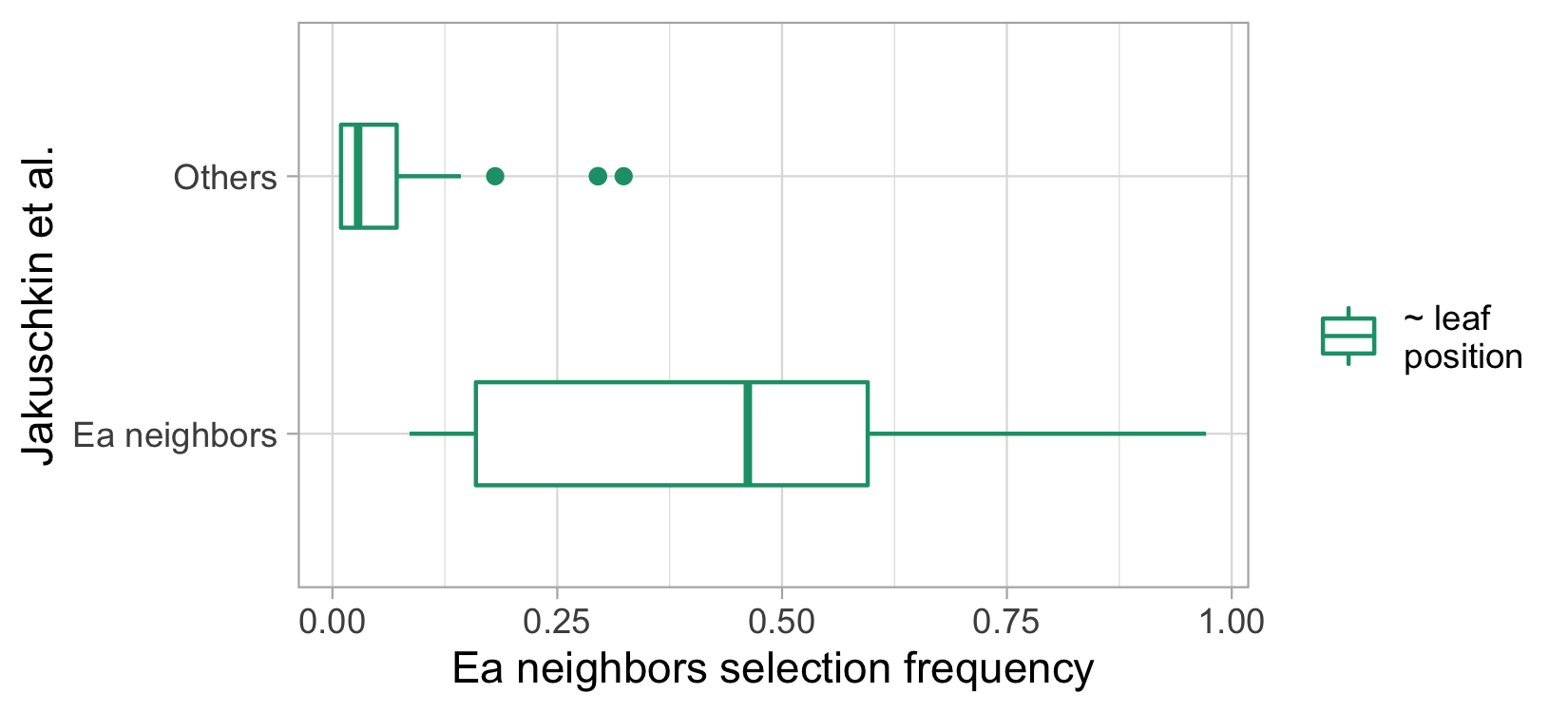}
    \caption{EMtree selection frequencies of pathogen neighbors compared to \citet{jakuch} results, computed on infected samples and adjusting for the leaf position (100 subs-samples). }
    \label{otujak}
\end{figure}

%  \begin{figure}[H]
%      \centering
%      \includegraphics[width=0.7\linewidth]{figs/propOTUS.png}
%      \caption{Proportions among EMtree results of OTUs common to both approaches.}
%      \label{compareOTUs}
%  \end{figure}
\modif{Using the dataset restricted to infected samples (39 observations for 114 OTUs) and correcting for the leaves position in the tree (proxy for their abiotic environment), \citet{jakuch} identifies a list of 26 OTUs likely to be directly interacting with the pathogen. Running EMtree on the same restricted dataset with the same correction yields a good concordance with  edge selection frequencies, as shown in Fig.~\ref{otujak}.}

%Fig.~\ref{otujak} compares this list to Ea neighbors selection frequencies obtained with EMtree run on infected samples only, a  restricted dataset of 39 samples and 114 species. The position of leaves in the tree controls their abiotic environment, so we can correct for the latter by including distances and orientation covariates in the model. 

%A first result is that all neighbors identified by \citet{jakuch} are also found as potential neighbors by EMtree (non-null selection frequencies). More precisely, Fig.~\ref{otujak} shows a nice separation of frequencies, with minimum frequency for "Ea neighbors" at about $0.09$ and a median frequency at about $0.03$ for "Others".

% For a sufficiently high threshold, it would be interesting to have a closer look at the  disagreements between the two studies. Table~\ref{comparjakusch} in Appendix \ref{app:results} details the outcomes comparison obtained with  a $90\%$ threshold for EMtree. This shows our approach finds four potential neighbors of Ea  which were not identified by \citet{jakuch}, among which three (F29, F32 and F579) are neighbors in the \textit{tree} model but not in the last model. This suggests that these connections are involved in the spatial spread of the pathogen in the oak tree. However, further investigations would be required to better understand the role of these unidentified OTUs. 

 \label{sec:illustration}

\section{Discussion} 
%\textcolor{gray}{To add in the discussion, about network comparison: {\sl We know that this variability exists across space \citep{PCM12},that it depends on abundances, but also traits, and local conditions \citep{PSG15}, and that different levels in networks are differently affected by local conditions, including the fact that species and interaction respond to different bioclimatic variables \citep{PGF17}.} }

The \modif{inference of species interaction network} is a challenging task, for which a series of methods have been proposed in the past years. Abundance data seems to be a promising source of information for this purpose. Here we adopt the formalism of graphical models to define a probabilistic model-based framework for the inference of such networks from abundance data.
Using a model-based approach offers several important advantages. First, it enables easy and explicit integration of environmental and experimental effects.  These could be modeled in a more flexible way using generalized additive models, which include non-linear effects \citep{hastie2017generalized}. 
Then, as it also relies on a formal statistical definition of a \textsl{species interaction network} in the context of graphical models, accounting for abiotic effects and modeling species interactions are two clearly defined and distinguished goals. Finally, all the underlying assumptions are explicitly stated in the model definition itself, and can therefore be discussed and criticized. \\

We developed an efficient method to infer sparse networks, which combines a multivariate Poisson mixed model for the joint distribution of abundances, with an averaging over all spanning trees to efficiently infer direct species interactions. As we do consider a mixture over all spanning trees, our approach remains flexible and can infer most types of statistical dependencies. An EM algorithm (EMtree) maximizes the likelihood of the result and returns each edge probability to be part of the network. An optional resampling step increases network robustness.

\modif{A simulation study in a heterogeneous environment  demonstrates  that EMtree  compares very well to alternative approaches. The proposed model can take all kind of covariates into account, which when ignored  can have  dramatic effects  on the inferred network structure, as showed here on empirical datasets.  Experiments on simulated data and illustrations also demonstrate that EMtree  computational cost remains very reasonable.}

\modif{Alternative methods used in this work all rely on an optimized threshold to tell an edge presence. This particular threshold is obtained after testing a grid of possible values which all yield a different network, and altogether build a path. Making this path available to the user is useful, as the final threshold might need modification and it gives the possibility to build edges scores  and get more than a binary result. We found few recent approaches doing this, which prevented us to study their performance in a way that did not impose a threshold.}\\

The proposed methodology could be extended in several ways.
\modif{Species abundances and interactions indeed vary across space, and depend on local conditions \citep{PCM12,PSG15}. This can either be considered as nuisance parameter or as feature of interest. In the first case, the method could be extended to account for the spatial autocorrelation of sampling sites, to obtain a "regional" interaction network corrected for this effect, i.e. assuming the network is the same in all sites. If of interest, variation across space and local conditions could be studied by comparing networks inferred from the different sampling locations. Networks comparison is a wide and interesting question and tools lack to check which edges are shared by a set of networks. The approach introduced by \citet{Loic} could be adapted to  EMtree framework.} Lastly,  It is also very likely that not all covariates nor even all species have been measured or observed. Another extension may therefore be to detect ignored covariates or missing species. To this purpose EMtree could probably be combined with the approach developed by \citet{genvieve} to identify missing actors. 

%Besides, ecological abundance data is usually collected in spatially organized sites.  Taking spatial autocorrelation would obviously improve the relevance of the proposed model. Since our model implies a latent structure, a natural way to take spatial dependencies into account would be to introduce a correlation between the latent vectors associated with each site, as done in multivariate geostatistical processes \citep{cressie1992statistics}. 

\paragraph{Data accessibility.}
The method developed in this paper is implemented in the R package EMtree available on GitHub: \url{https://github.com/Rmomal/EMtree}.

\paragraph{Acknowledgement.}
The authors thank P. Gloaguen and M. Authier for helpful discussions and C. Vacher for providing the oak data set. This work is partly funded by ANR-17-CE32-0011 NGB and by a public grant as part of the Investissement d'avenir project, reference ANR-11-LABX-0056-LMH, LabEx LMH.

\paragraph{Author's contributions statement.}

All authors conceived the ideas and designed methodology; R. Momal developed and tested the algorithm. All authors led the writing of the manuscript, contributed critically to the drafts and gave final approval for publication.

%%%%%%%%%%%%%%%%%%%%%%%%%%%%%%%%%%%%%%%%%%%%%
%\bibliographystyle{science}
\bibliographystyle{apsrev} %tested plainnat
\bibliography{biblio.bib}

\begin{appendix}
\newpage
\resetlinenumber
\section{Supplement: Methods}
\subsection{Variational EM in the observed layer} \label{app:VEM} \modif{\paragraph{A reminder on EM and VEM.} Expectation-Maximisation \citep[EM:][]{DLR77} has become the standard algorithm for the maximum likelihood inference of latent variable models. Denoting $\gammab$ the unknown parameter, $\Yb$ the observed variables and $\Hb$ the latent variables, the aim of EM is to maximise the {\sl observed} (log-)likelihood $\log p_\gammab(\Yb)$. 
In the model defined in Section \ref{sec:model}, the set of parameter to estimate is $\gammab = (\betab, \Sigmab, \thetab)$ and the latent variables are $\Hb = (\Zb, T)$.
Because the {\sl complete} (log-)likelihood $\log p_\gammab(\Yb, \Hb)$ is often much easier to handle, EM alternatively evaluates the conditional distribution of the latent variables $p_\gammab(\Hb \mid \Yb)$ (E step) and updates the parameter estimates by maximizing the conditional expectation of the complete log-likelihood (M step).}

\modif{Unfortunately, for many models, the conditional distribution $p_\gammab(\Hb \mid \Yb)$ is intractable. The variational EM (VEM) algorithm has been designed to deal with such cases. Briefly speaking, the E step (during which the intractable conditional distribution should be evaluated) is replaced with a VE step, during which an approximate distribution $\pt(\Yb) \simeq p_\gammab(\Hb \mid \Yb)$ is determined. Actually, the VEM algorithm maximizes a lower bound of the genuine log-likelihood, similar to this given in Eq.~\eqref{eq:LowerBound} \citep[see][for an introduction]{OrW10,BKM17}.}

\paragraph{Application to the Poisson log-normal model.} 
To estimate the fixed regression parameters gathered in $\thetab$, we resort to a surrogate model where the entries of the abundance matrix $\Yb$ still have the conditional distribution given in Eq.~\eqref{eq:pY.Z}, but where the distribution of the $\Zb_i$ is not constrained to be faithful to a specific graphical model. Namely, the latent vectors $\Zb_i$ are only supposed to be independent and identically distributed (iid) Gaussian with distribution $\Ncal(0, \Sigmab)$, without any restriction on $\Sigmab$. 

This surrogate model is actually a Poisson log-normal model as introduced by \cite{AiH89}, the parameters of which can be estimated using a variational approximation similar to this introduced in \cite{CMR18}. 
%Namely, instead of maximizing the log-likelihood $\log p(\Yb)$ with respect to the parameters $\thetab$ and $\Sigmab$ using a regular EM algorithm, we maximize the lower bound of it
\modif{More specifically, we maximize with respect to the parameters $\thetab$ and $\Sigmab$ the following lower bound of the log-likelihood $\log p(\Yb)$:}
\begin{linenomath} 
\begin{equation}\label{eq:LowerBound}
\mathcal{J}(\Yb; \thetab, \Sigmab, \pt) := \log p_{\thetab, \Sigmab}(\Yb) - KL\left(\pt(\Zb) || p_{\thetab, \Sigmab}(\Zb \mid \Yb)\right),
\end{equation}
\end{linenomath}
where $KL(q||p)$ stands for Küllback-Leibler divergence between distributions $q$ and $p$ and where the approximate distribution $\pt(\Zb)$ 
%(which approximates $\pt_{\thetab, \Sigmab}(\Zb \mid \Yb)$) 
is chosen to be Gaussian. This means that each conditional distribution $p(\Zb_i \mid \Yb_i)$ is approximated with a normal distribution $\Ncal(\mbt_i, \Sbt_i)$. As shown in \cite{CMR18}, $\mathcal{J}(\Yb, \thetab, \Sigmab, \pt)$ is bi-concave in $(\thetab, \Sigmab)$ and $\{(\mbt_i, \Sbt_i)_i\}$, so that gradient ascent can be used. The {\tt PLNmodels} R-package --available on GitHub-- provides an efficient implementation of it.

The entries of the $\mbt_i$ and $\Sbt_i$ provide us with approximations of the conditional expectation,  variance and covariance of the $Z_{ij}$ conditionally on the $\Yb$, which we use to get the estimates $\widehat{\sigma}_j^2$ and $\widehat{\rho}_{jk}$ given in Eq.~\eqref{eq:sigma.rho}. More specifically, we use $\Esp(Z_{ij} \mid \Yb_i) \simeq \mt_{ij}$, $\Esp(Z^2_{ij} \mid \Yb_i) \simeq \mt_{ij}^2 + \St_{i,jj}$ and $\Esp(Z_{ij} Z_{ik} \mid \Yb_i) \simeq \mt_{ij} \mt_{ik} + \St_{i, jk}$.
\subsection{EM in the latent layer} \label{app:EM} %%%%%%%%%%%%%%%%%%%%%%%%%%%%%%%%%%%%%%%%%%%%%%%%%%%%%%%%%%%%%%%%%%%%%%%%%%%%%%%%%%%%%%%%%%%%
\subsubsection*{Complete log-likelihood conditional expectation}
%%%%%%%%%%%%%%%%%%%%%%%%%%%%%%%%%%%%%%%%%%%%%%%%%%%%%%%%%%%%%%%%%%%%%%%%%%%%%%%%%%%%%%%%%%%%
Because of the specific form given in Eq.~\eqref{eq:pZfact}, and because the $\Zb_i\mid T$  are Gaussian, we have that
\begin{linenomath*} 
\begin{align}
    \label{eq:logpZ.T}
    \log{p_\Sigmab}(\Zb \mid T)
    & = \sum_{j=1}^p \sum_{i=1}^n \log P(Z_{ij} \mid T) 
    + \sum_{(j, k) \in T} \sum_{i=1}^n \log \left(\frac{P(Z_{ij}, Z_{ik})}{P(Z_{ij})P(Z_{ik})}\right) \nonumber \\
    & = - \frac{n}2 \log \sigma_j^2 -\frac12 \sum_{j=1}^p \sum_{i=1}^n \frac{Z_{ij}^2}{\sigma_j^2}
    - \frac{n}2 \sum_{(j, k) \in T} \log (1- \rho_{jk}^2) \\
    & \quad - \frac12 \sum_{(j, k) \in T} \frac1{1- \rho_{jk}^2} \sum_{i=1}^n \left(
    \rho^2_{jk} \frac{Z_{ij}^2}{\sigma_j^2} + \rho^2_{jk} \frac{Z_{ik}^2}{\sigma_k^2} - 2\rho_{jk} \frac{Z_{ij} Z_{ik}}{\sigma_j \sigma_k}
    \right)
    + \text{cst} \nonumber 
\end{align}
\end{linenomath*}

where the constant term does not depend on any unknown parameter. 
In the EM algorithm, we have to maximize the conditional expectation of Eq.~\eqref{eq:logpZ.T} with respect to the variances $\sigma_j^2$ and the correlation coefficients $\rho_{jk}$. The resulting estimates take the usual forms, but with the conditional moments of the $Z_{ij}$, that is
\begin{linenomath*}
\begin{equation} \label{eq:sigma.rho}
\widehat{\sigma}_j^2  = \frac1n \sum_i \Esp(Z^2_{ij}  \mid  \Yb),
\qquad
\widehat{\rho}_{jk}   = \frac1n \sum_i \Esp(Z_{ij} Z_{ik}  \mid  \Yb) \left/ (\widehat{\sigma}_j \widehat{\sigma}_k) \right. .
\end{equation}
\end{linenomath*}
which do not depend on T. 
The maximized conditional expectation of Eq.~\eqref{eq:logpZ.T} becomes
\begin{linenomath*}
\begin{equation} \label{eq:ElogpZ.T.Y}
    \Esp\left(\log p_{\widehat{\Sigmab}}(\Zb \mid T) \mid \Yb\right)
    = - \frac{n}2 \log \widehat{\sigma}_j^2 
    - \frac{n}2 \sum_{(j, k) \in T} \log (1- \widehat{\rho}_{jk}^2)
    + \text{cst}.
\end{equation}
\end{linenomath*}
We are left with the writing of the conditional expectation of the first two terms of the logarithm of Eq.~\eqref{eq:PTZY}, once optimized in $\Sigmab$. Combining Eq.~\eqref{eq:pT} and Eq.~\eqref{eq:ElogpZ.T.Y}, and noticing that the probability for an edge to be part of the graph is the sum of the probability of all the trees than contain this edge, we get (denoting $\log \widehat{\psi}_{jk} = (1- \widehat{\rho}_{jk}^2)^{-n/2}$)
\begin{linenomath*}
\begin{align*}
    \Esp\left(\log p_\betab (T) + \log p_{\widehat{\Sigmab}}(\Zb \mid T) \mid \Yb\right)
    & = \sum_{T \in \Tcal} p(T \mid \Yb) \left(\log p_\betab(T) + \log p_{\widehat{\Sigmab}}(\Zb \mid T) \right) \\
    & = -\log B + \sum_{T \in \Tcal} p(T \mid \Yb) \sum_{(j, k) \in T} \left(\log \beta_{jk} +\log \widehat{\psi}_{jk} \right) + \text{cst} \\
    & = -\log B + \sum_{(j, k)} \prob\{(j, k) \in T \mid \Yb\} \left(\log \beta_{jk} + \log \widehat{\psi}_{jk} \right) + \text{cst},
\end{align*}
\end{linenomath*}
which gives Eq.~\eqref{expectation}.\\
 
As explained in the section above, we approximate expectations and probabilities conditional on $\Yb$ by their variational approximation.
%\textcolor{red}{
%Because $T$ is independent of $Y$ given $Z$, we have that
%$$
%\log p(T \mid Y) = \log  \e_{Z \mid Y} \left(p(T, Z)\right), 
%$$
%the variational approximation of which is 
%$
%\log \et \left(p(T, Z)\right)
%\simeq \et \left(\log p(T, Z)\right)
%$,
%so we define
%$$
%\log \pt(T \mid Y)
%:= \et \left(\log p(T, Z)\right)
%= \sum_{j, k \in T} \log (\beta_{jk} \psi_{jk}) + \text{cst}
%$$
%(where the constant term does not depend on $T$), that is
%$
%\pt(T \mid Y) 
%= \left. \prod_{j, k \in T} \beta_{jk} \psi_{jk} \right/ C,
%$
%where $C$ is the normalizing constant:
%$C = \sum_T \prod_{j, k \in T} \beta_{jk} \psi_{jk}$.
%}
This provides us with the approximate conditional distribution of the tree $T$ given the data $\Yb$:
\begin{linenomath*}
$$
\pt(T  \mid  \Yb) = \left. \prod_{jk \in T} \beta_{jk} \widehat{\psi}_{jk}  \right/ C,
$$
\end{linenomath*}
where $C$ is the normalizing constant: $C = \sum_T \prod_{j, k \in T} \beta_{jk} \widehat{\psi}_{jk}$. The intuition behind this approximation is the following: according to Eq. \eqref{eq:pT}, the marginal probability a tree $T$ is proportional to the product of the weights $\beta_{jk}$ of its edges. The conditional distribution probability of tree is proportional to the same product, the weights $\beta_{jk}$ being updated as $\beta_{jk} \widehat{\psi}_{jk}$, where $\widehat{\psi}_{jk}$ summarizes the information brought by the data about the edge ($j, k$).

%%%%%%%%%%%%%%%%%%%%%%%%%%%%%%%%%%%%%%%%%%%%%%%%%%%%%%%%%%%%%%%%%%%%%%%%%%%%%%%%%%%%%%%%%%%%
\subsubsection*{Steps E and M}
%%%%%%%%%%%%%%%%%%%%%%%%%%%%%%%%%%%%%%%%%%%%%%%%%%%%%%%%%%%%%%%%%%%%%%%%%%%%%%%%%%%%%%%%%%%%
\begin{description}
 \item[E step:]
From the above computation we get the following approximation:
 
\begin{linenomath*}
$$\mathds{P}(\{j,k\}\in T  \mid \Yb) \simeq 1 - \sum_{T:jk\notin T} \pt(T \mid \Yb),$$
\end{linenomath*}
and so we define $p_{jk}$ as follows: \begin{linenomath*}
$$P_{jk}= 1 - \frac{\sum_{T:jk\notin T} \prod_{j, k \in T} \beta_{jk} \psi_{jk}}{\sum_T \prod_{j, k \in T} \beta_{jk} \psi_{jk}}.$$
 \end{linenomath*}
 $P_{jk}$ can be computed with Theorem \ref{thm:MTT}, letting $[\Wb^h]_\jk = \beta^h_\jk \widehat{\psi}_\jk$ and $\Wb^h_{\setminus \jk} = \Wb^h$ except for the entries $(j, k)$ and $(k, j)$ which are set to 0. The modification of $\Wb^h_{\setminus \jk}$ with respect to $\Wb^h$ amounts to set to zero the weight product, and so the probability, for any tree $T$ containing the edge $(j, k)$. As a consequence, we get
\begin{linenomath*} 
$$
 P^{h+1}_\jk = 1 - \left|Q_{uv}^*(\Wb^h_{\setminus \jk})\right| \left/ \left|Q_{uv}^*(\Wb^h)\right| \right..
 $$
 \end{linenomath*}
 \item[M step:] Applying Lemma \ref{lem:Meila} to the weight matrix $\betab$, the derivative of $B$ with respect to $\beta_\jk$ is 
\begin{linenomath*}
$$
 \partial_{\beta_\jk} B = [\Mb(\betab)]_\jk \times B
 $$
 \end{linenomath*}
 then the derivative of \eqref{expectation} with respect to $\beta_\jk$ is null for
 $\beta^{h+1}_\jk = P_\jk^{h+1} \left/ [\Mb(\betab^h)]_\jk \right.$.
\end{description}

\subsection{Matrix tree theorem}\label{app:MTT} For any matrix $\Wb$, we denote its entry in row $u$ and column $v$ by $[\Wb]_{uv}$. We define the Laplacian matrix $\Qb$ of a symmetric matrix $\Wb=[w_\jk ]_{1\leq j,k\leq p}$ as follows :
\begin{linenomath*}
\[
 [\Qb]_\jk =\begin{cases}
    -w_\jk  & 1\leq j<k \leq p\\
    \sum_{u=1}^p w_{ju} & 1\leq j=k \leq p.
    \end{cases}
\]
\end{linenomath*}
We further denote $\Wb^{uv}$ the matrix $\Wb$ deprived from its $u$th row and $v$th column and we remind that the $(u, v)$-minor of $\Wb$ is the determinant of this deprived matrix, that is $|\Wb^{uv}|$.

\begin{theorem}[Matrix Tree Theorem  \cite{matrixtree,MeilaJaak}] \label{thm:MTT}
    For any symmetric weight matrix W, the sum over all spanning trees of the product of the weights of their edges is equal to any minor of its Laplacian. That is, for any $1 \leq u, v \leq p$,
   \begin{linenomath*}
   \[
    W := \sum_{T\in\mathcal{T}} \prod_{(j, k)\in T} w_\jk  = |\Qb^{uv}|.
    \]
    \end{linenomath*}
\end{theorem}    

In the following, without loss of generality, we will choose $\Qb^{pp}$. As an extension of this result, \cite{MeilaJaak} provide a close form expression for the derivative of $W$ with respect to each entry of $\Wb$. 

\begin{lemma} [\cite{MeilaJaak}] \label{lem:Meila}
    Define the entries of the symmetric matrix $\Mb$ as
\begin{linenomath*}
\[    
 [\Mb]_\jk =\begin{cases}
    \left[(\Qb^{pp})^{-1}\right]_{jj} + \left[(\Qb^{pp})^{-1}\right]_{kk} -2\left[(\Qb^{pp})^{-1}\right]_\jk & 1\leq j<k < p\\
    \left[(\Qb^{pp})^{-1}\right]_{jj} & k=p, 1\leq j \leq p  \\
    0 & 1\leq j=k \leq p.
    \end{cases}
\]
\end{linenomath*}
it holds that
\begin{linenomath*}
$$
\partial_{w_\jk} W = [\Mb]_\jk  \times W.
$$
\end{linenomath*}
\end{lemma}

\section{Supplement}
\label{app:results} \subsection{Simulation results}

\subsubsection{Effect of dataset dimensions}
\begin{table}[ht]
\centering
\begin{tabular}{l|l|rrrrrr}
\multicolumn{2}{l|}{} & \multicolumn{1}{c}{SpiecEasi} & \multicolumn{1}{c}{gCoda} & \multicolumn{1}{c}{ecoCopula} & \multicolumn{1}{c}{MRFcov} & \multicolumn{1}{c}{MInt} & \multicolumn{1}{c}{EMtree} \\ 
\hline
\multirow{3}{*}{{\rotatebox[origin=c]{90}{Easy}}} 
    & Cluster & 0.86  (0.20) & 0  (0.08) & 0.33  (0.14) & 0.74  (0.06) & 0.38  (0.17) & 0.12  (0.09) \\ 
    & Erdös   & 0.86  (0.21) & 0  (0.15) & 0.29  (0.15) & 0.73  (0.05) & 0.38  (0.15) & 0.12  (0.08) \\ 
    & Scale-free & 0.92  (0.04) & 0  (0.04) & 0.33  (0.11) & 0.88  (0.02) & 0.73  (0.09) & 0.34  (0.08) \\  \hline
\multirow{3}{*}{{\rotatebox[origin=c]{90}{Hard}}} & Cluster  &0.87  (0.12) & 0  (0.20) & 0.15  (0.18) & 0.78  (0.05) & 0.77  (0.09) & 0.36  (0.09) \\ 
 & Erdös.     & 0.88  (0.11) & 0  (0.24) & 0  (0.15) & 0.78  (0.05) & 0.77  (0.10) & 0.35  (0.09) \\ 
 & Scale-free & 0.94  (0.05) & 0  (0.13) & 0  (0.16) & 0.89  (0.03) & 0.94  (0.03) & 0.56  (0.07) \\ \hline
\end{tabular}
\caption{Medians and standard-deviation of FDR computed on 100 graphs of each type (\textit{easy}: $n=100, p=20$, \textit{hard}: $n=50, p=30$)}
\label{medFDR}
\end{table}

\begin{table}[ht]
\centering
\begin{tabular}{r|l|lllllll}
\multicolumn{2}{l|}{} & \multicolumn{1}{c}{SpiecEasi} & \multicolumn{1}{c}{gCoda} & \multicolumn{1}{c}{ecoCopula} & \multicolumn{1}{c}{MRFcov} & \multicolumn{1}{c}{MInt} & \multicolumn{1}{c}{EMtree} \\ \hline
\multirow{3}{*}{{\rotatebox[origin=c]{90}{Easy}}} &Cluster & 0.16  (0.11) & 0.05  (0.07) & 1.04  (0.48) & 2.26  (0.58) & 0.30  (0.13) & 0.62  (0.14) \\ 
& Erdös & 0.15  (0.09) & 0.06  (0.08) & 0.95  (0.50) &2.23  (0.42) & 0.30  (0.14) & 0.58  (0.12) \\ 
 & Scale-free & 0.63  (0.13) & 0.08  (0.07) & 0.92  (0.30) &4.86  (0.44) & 0.81  (0.25) & 0.96  (0.08) \\ \hline
\multirow{3}{*}{{\rotatebox[origin=c]{90}{Hard}}}  & Cluster & 0.21  (0.08) & 0.02  (0.03) & 0.02  (0.17) & 1.65  (0.33) & 0.68  (0.30) & 0.43  (0.10) \\ 
 & Erdös & 0.21  (0.08) & 0.02  (0.02) & 0.00  (0.18) & 1.56  (0.32)& 0.66  (0.25) & 0.42  (0.10) \\ 
 & Scale-free & 0.61  (0.12) & 0.04  (0.03) & 0.08  (0.24) &3.29  (0.40) & 3.63  (1.08) & 0.94  (0.09) \\ 
   \hline
\end{tabular}

\caption{Medians and standard-deviation of density ratio computed on 100 graphs of each type (\textit{easy}: $n=100, p=20$, \textit{hard}: $n=50, p=30$)}
\label{meddens}
\end{table}

\begin{table}[ht]
\centering
\begin{tabular}{l|l|rrrrrr}
\multicolumn{2}{l|}{} & \multicolumn{1}{c}{SpiecEasi} & \multicolumn{1}{c}{gCoda} & \multicolumn{1}{c}{ecoCopula} & \multicolumn{1}{c}{MRFcov} & \multicolumn{1}{c}{MInt} & \multicolumn{1}{c}{EMtree} \\ \hline
\multirow{3}{*}{{\rotatebox[origin=c]{90}{Easy}}} & Cluster & 1.77 & 13.89 & 1.74 & 0.00 & 0.00 & 0.00 \\
 & Erdös & 0.68 & 11.95 & 0.99 & 0.00 & 0.83 & 0.00 \\
 & Scale-free & 0.00 & 1.88 & 0.00 & 0.00 & 0.00 & 0.00 \\ \hline
\multirow{3}{*}{{\rotatebox[origin=c]{90}{Hard}}} & Cluster & 0.00 & 14.05 & 23.40 & 0.00 & 0.00 & 0.00 \\
 & Erdös & 0.00 & 20.85 & 27.28 & 0.00 & 0.00 & 0.00 \\
 & Scale-free & 0.00 & 5.97 & 15.46 & 0.00 & 0.00 & 0.00 \\ \hline
\end{tabular}
\caption{Percentage of empty networks computed on 100 graphs of each type (\textit{easy}: $n=100, p=20$, \textit{hard}: $n=50, p=30$)}
\label{empty}
\end{table}

\begin{table}[ht]
\centering
\begin{tabular}{lrrrrrr}
 & \multicolumn{1}{c}{SpiecEasi} & \multicolumn{1}{c}{gCoda} & \multicolumn{1}{c}{ecoCopula} & \multicolumn{1}{c}{MRFcov} & \multicolumn{1}{c}{MInt} & \multicolumn{1}{c}{EMtree} \\ 
  \hline
Easy & 29.73  (2.00) & ~1.29  (~0.30) & 28.14  (1.46) & 48.7  (2.32) & 138.13  (39.60) & 50.19  (7.81) \\ 
  Hard & 29.38  (1.31) & 40.73  (20.94) & 27.16  (1.30) & 14.1  (0.36) & ~95.27  (46.34) & 23.59  (2.09) \\ 
   \hline
\end{tabular}
\caption{Median and standard-deviation of running times in seconds  for scale-free structures, for two dataset dimensions (\textit{easy}: $n=100$, $p=20$, \textit{hard}: $n=50$, $p=30$). }
\label{timeSF}
\end{table}

\begin{figure}[H]
    \centering
    \includegraphics[width=10cm]{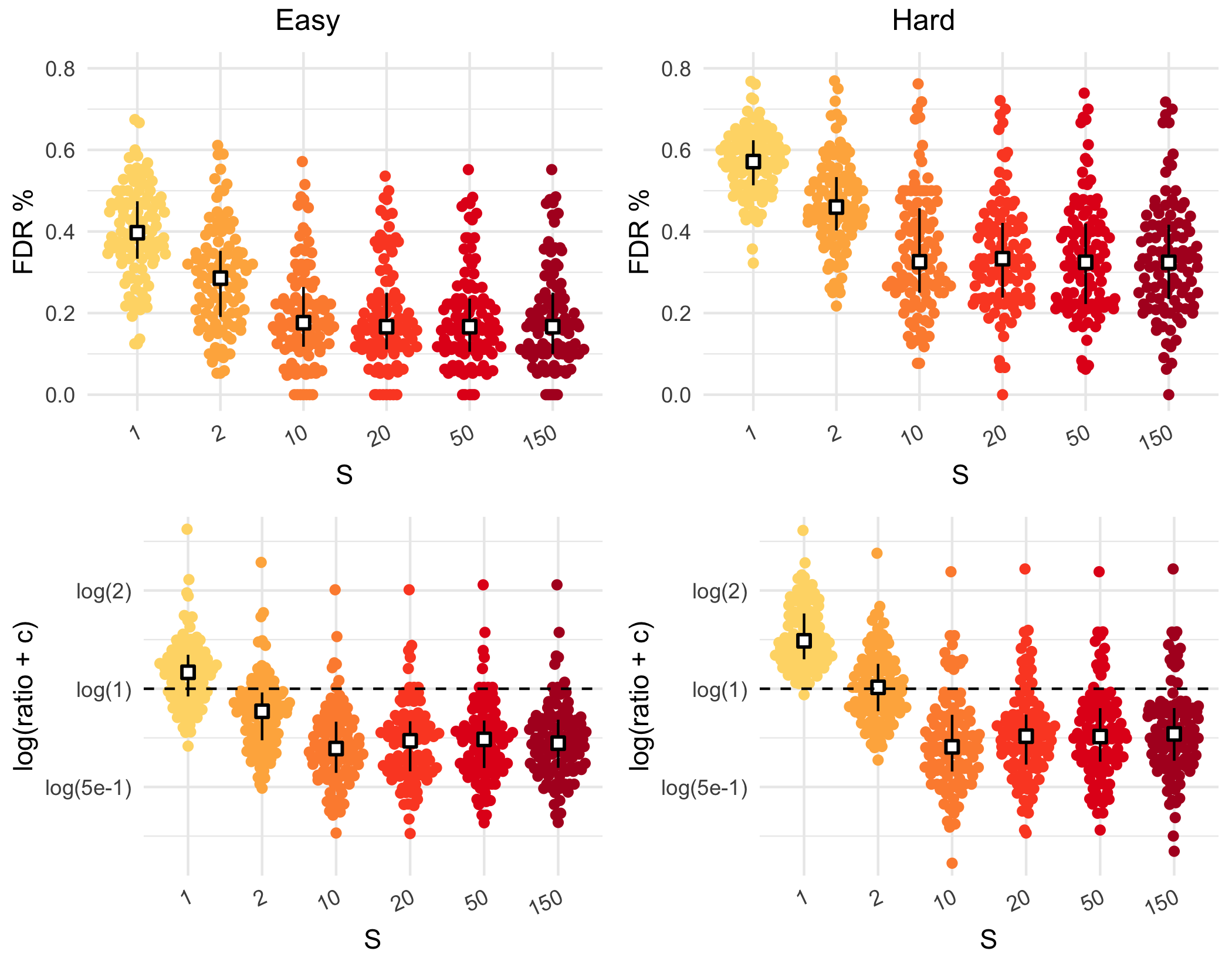}
    \caption{FDR and density ratio measures of EMtree with varying values of number of sub-samples $S$ (Erdös structure).}
    \label{Seffect}
\end{figure}
\begin{table}[ht]
\centering
\begin{tabular}{l|rrrrrr}
  $S$ & \multicolumn{1}{c}{1} & \multicolumn{1}{c}{2} & \multicolumn{1}{c}{10} & \multicolumn{1}{c}{20} & \multicolumn{1}{c}{50} & \multicolumn{1}{c}{150} \\ \hline
  Easy & 0.66  (0.15) & 1.86  (0.23) & 7.00  (0.81) & 12.29  (1.27) & 29.50  (3.39) & 87.30  (10.36) \\ 
  Hard & 0.45  (0.12) & 1.44  (0.14) & 5.06  (0.78) & 8.97  (0.87) & 23.35  (2.40) & 69.29  (10.83) \\ 
   \hline
\end{tabular}
\caption{Median and standard-deviation running-time values in seconds for inference of Erdös structure with EMtree and different values of the number of sub-samples $S$.}
\label{timesS}
\end{table}

\subsubsection{Effect of network density}

\begin{table}[H]
\centering
\begin{tabular}{l|rr|rr}
    & \multicolumn{1}{c}{$n < 50$} & \multicolumn{1}{c}{$n\geq 50$} & \multicolumn{1}{c}{$p < 20$} & \multicolumn{1}{c}{$p\geq 20$} \\  \hline
  EMtree    &   0.41 (0.11)	&   0.60 (0.15) &   0.38 (0.12) &    0.71 (0.21)      \\ 
  gCoda     &   0.12 (0.47)	&   0.07 (0.03) &   0.05 (0.03) &    0.09 (0.06)     \\ 
  SpiecEasi &   2.41 (0.25)	&   2.41 (0.25) &   2.39 (0.25) &    2.42 (0.25)      \\ 
   \hline
\end{tabular}
\caption{Median and standard-deviation of running times for each method in seconds, for $n$ and $p$ parameters. corresponding to Erdös and cluster structures with $5/p$ densities.}
\label{timeDenser}
\end{table}

\subsection{Illustrations}
\subsubsection{Effect of the edge frequency threshold}

\begin{figure}[H]
    \centering
    \includegraphics[width=\linewidth]{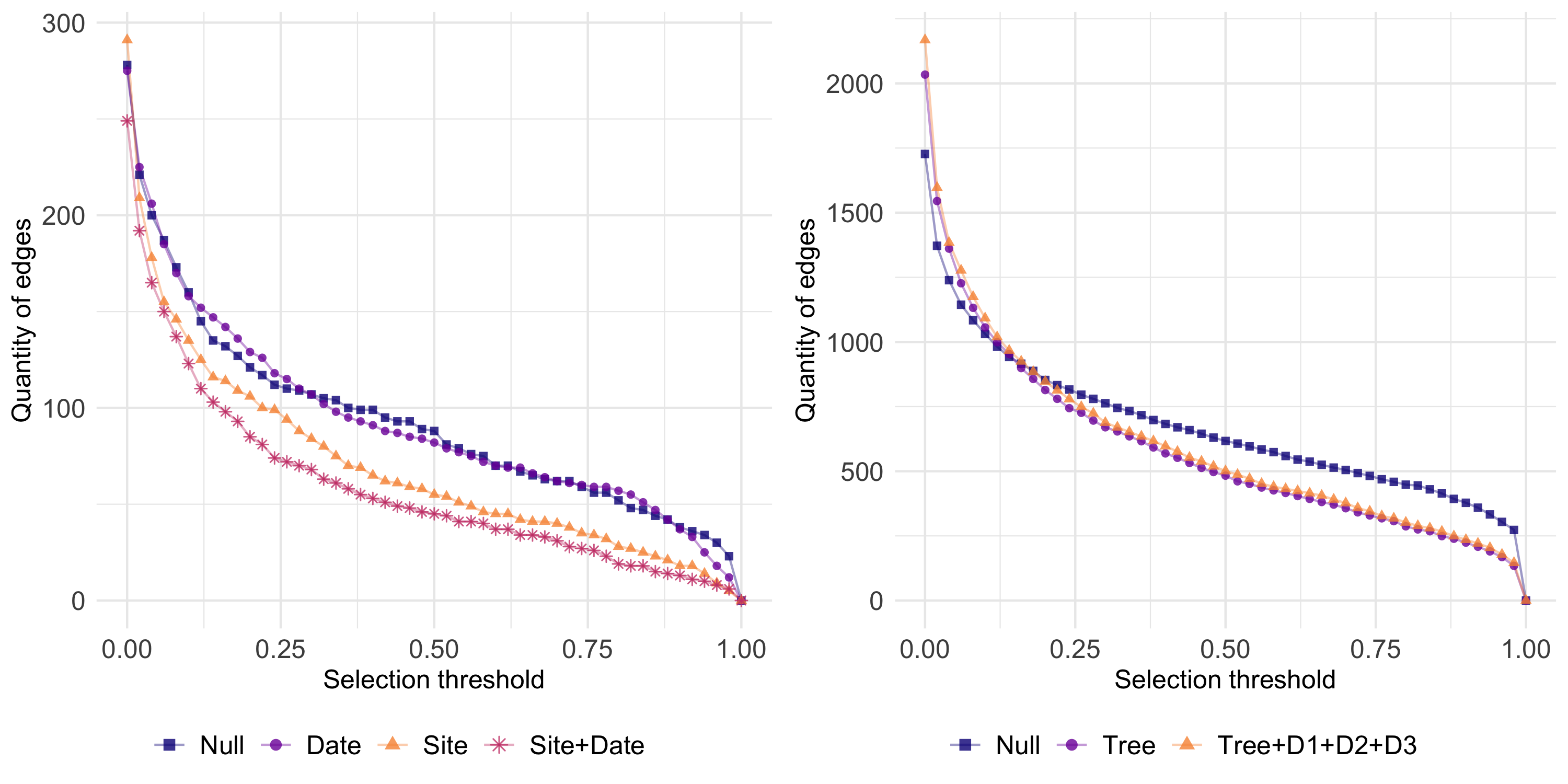}
    \caption{Quantity of selected edges as a function of the selection threshold (\textit{left}: Fatala fishes, \textit{right}: oak mildew.)}
    \label{QETOak}
\end{figure}

The curves displayed on Fig. \ref{QETOak} are very smooth, which illustrates the difficulty of setting this threshold.

\subsubsection{Fatala River fishes}
\label{names_Baran}
\paragraph{Species names with highest betweenness scores:}
  13: Galeoides decadactylus, 17:Ilisha africana,19: Liza grandisquamis, 22:  Pseudotolithus brachygnatus, 25: Pellonula leonensis, 30: Pseudotolithus typus.

\end{appendix}

\end{document}